\documentclass[preprint]{aastex}

\def\Hthreep{H$_3^+$}          % H3+
\def\NtwoHp{N$_2$H$^+$}        % N2H+ 
\def\CeiO{C$^{18}$O}        % N2H+ 
\def\CsvO{C$^{17}$O}        % N2H+ 
\def\CtfS{C$^{34}$S}        % N2H+ 
\def\Ntwo{N$_2$}               % N2
\def\av{$A_V$}               % N2
\newcommand{\nhtwo}{\mbox{n$_{H_{2}}$}}
\newcommand{\cc}{\mbox{cm$^{-3}$}}

\begin{document}

%\title{Depletion of CO \& CS and the Production of
%N$_2$H+ in the Dark Cloud Associated with IC 5146}
\title{Molecular Excitation and Differential Gas-Phase Depletions in
the IC 5146 Dark Cloud}

\author{Edwin A. Bergin}
\affil{Harvard-Smithsonian Center for Astrophysics, 60 Garden Street,
Cambridge, MA 02138} \email{ebergin@cfa.harvard.edu}

\author{David R. Ciardi}
\affil{Department of Astronomy, University of Florida, Gainesville, FL
32611} \email{ciardi@astro.ufl.edu}

\author{Charles J. Lada}
\affil{Harvard-Smithsonian Center for Astrophysics, 60 Garden Street,
Cambridge, MA 02138} \email{clada@cfa.harvard.edu}

\author{Joao Alves}
\affil{European Southern Observatory, Karl-Schwarzschild-Strasse 2, 85748
Garching, Germany}
\email{jalves@eso.org}

\author{Elizabeth A. Lada}
\affil{Department of Astronomy, University of Florida, Gainesville, FL
32611} \email{lada@astro.ufl.edu}

\slugcomment{Accepted by The Astrophysical Journal}
\begin{abstract}
We present a combined near-infrared and molecular-line study of 
25$' \times 8'$ area in the Northern streamer of the IC 5146 cloud.  
Using the technique pioneered
by \citet{lada94}, we construct a Gaussian smoothed map of the 
infrared extinction with the same resolution as the molecular line observations
in order to examine correlations of integrated intensities
and molecular abundances with extinction for \CsvO ,
\CtfS, and \NtwoHp .
We find that over a visual extinction range of 0 to 40 magnitudes,
there is good evidence 
for the presence of differential gas-phase depletions in
the densest portions of IC 5146.  Both CO and CS exhibit a statistically
significant (factor of $\sim 3$) abundance reduction 
near \av\ $\sim 12$ magnitudes while, in direct contrast,
at the highest extinctions, \av $> 10$ magnitudes,
\NtwoHp appears relatively undepleted. 
Moreover, for \av\ $< 4$ magnitudes there exists little or no \NtwoHp .
This pattern of depletions is 
consistent with the predictions of chemical theory.
Through the use of a time and depth dependent
chemical model we show that the near-uniform or rising \NtwoHp\ abundance with
extinction is a direct result of a reduction in 
its destruction rate at high extinction due to the
predicted and observed depletion of CO molecules.  
The observed abundance threshold for \NtwoHp , $A_V^{th} \sim 4$ mag, is examined
in the context of this same model and we demonstrate how this technique can
be used to test the predictions of depth-dependent chemical models. 
Finally, we find that cloud density gradients can have a significant effect
on the excitation and detectability of high dipole moment molecules,
which are typically far from local thermodynamic equilibrium.  Density 
gradients also cause chemical changes as reaction rates and 
depletion timescales are density dependent. Accounting 
for such density/excitation gradients is crucial to a correct determination 
and proper interpretation of molecular abundances.

\end{abstract}


\section{Introduction}

The determination of masses, densities, energetics, and chemistry for cold
molecular cores is of fundamental importance for our understanding of how
and why stars form.   Unfortunately, the dominant molecule in interstellar
molecular clouds, H$_2$, is generally unobservable at the low temperatures
associated with star forming cores.  Consequently, the most fundamental
properties of molecular clouds (and the embedded star-forming cores) have
been determined almost exclusively from observations of rarer trace
molecules (and their isotopes) such as CO, CS, and NH$_3$.  However, these
secondary molecular tracers have been calibrated only in a handful of
sources, and because of variations in chemical abundances, chemical
evolution, and excitation conditions, application to other sources is
inherently uncertain, making accurate determination of important physical
properties (e.g., sizes, masses, and energetics) of molecular clouds not
always easy or possible.

Because of the apparent constancy of the gas-to-dust ratio in molecular
clouds \citep{js74, bsd78}, the most direct method to trace the hydrogen
content of a molecular cloud is to determine the distribution of dust
throughout the cloud.  In principle, the distribution of the dust can be
derived from far-infrared and millimeter wavelength dust emission, but the
grain opacity at high densities is poorly constrained, limiting the
accuracy of the dust (and hence, H$_2$) column density determination
\citep{kram98}. Recently, \citet{lada94} developed a powerful technique for
mapping the large scale distribution of dust using multi--wavelength
near-infrared imaging. By measuring the near-infrared color excess of
stars behind a cloud the line of sight dust extinction (and hence, total
column density) can be directly determined.  
The technique enables measurements of the dust
distribution over a significant of range of angular scales and extinction
\citep{crd98, lal99}.  Because the near-infrared extinction law does not
vary significantly with grain growth in cold cores \citep{math90}, high
extinction observed towards cloud cores can be anchored to the
well-calibrated H$_2$-to-dust ratio at lower extinction.

When the near-infrared excess data is combined with molecular gas
emission-line observations, direct determination of the molecular
abundances relative to H$_2$ can be obtained; thus, allowing a more
accurate determination of the physical and chemical properties of
molecular clouds and their embedded cores \citep{all99}.  Indeed,
$^{12}$C$^{18}$O, a molecule often assumed to trace accurately the distribution
of material in cold cores, has been found to show evidence for a reduction
in abundance at high extinction which has been attributed to the freezing
of gas-phase CO onto the surfaces of cold dust grains (i.e., depletion)
in the cores of IC 5146 \citep{kram99} and L 977
\citep{all99}.

\citet{kram99} found compelling evidence for depletion of CO on to dust grains in a
cold core located in the IC 5146 dark cloud.  The finding was based
primarily upon the relationship between the integrated intensity of
C$^{18}$O and visual extinction derived from near-infrared color excess
measurements. The C$^{18}$O integrated intensity was found to be
well-correlated with $A_V$ for $A_V \lesssim 10$ mag.  For $A_V \gtrsim
10$ mag, the relationship flattens out, and the C$^{18}$O integrated
intensity is nearly constant out to a visual extinction of $A_V \sim 30$
mag.  C$^{17}$O was used to check for opacity effects, and the C$^{18}$O
was found to be optically thin throughout the region.  \citet{kram99}
concluded that the most likely explanation was the depletion of CO on to
the surface of dust grains. Thus, masses, and densities of cold dark cores
in IC5146 derived from $^{12}$C$^{18}$O may underestimate the true mass and
densities by as much as 30\%.

However, the study by \citet{kram99} covered a very small region ($3\farcm
3 \times 3\farcm 3$) and was limited to a single core in the IC 5146 dark cloud.
In this paper, we present a more detailed study of the molecular gas
abundances over a much larger region in the IC 5146 dark cloud ($25\arcmin
\times 8\arcmin$) at moderate resolution (50\arcsec). Our primary goal is
to perform, for a variety of molecular species, a direct comparison of the
molecular emission and the line of sight dust column density in the dark
cloud associated with the young cluster IC 5146.  By using the line of
sight dust extinction to trace the H$_2$ column density and comparing that
to the measured molecular emission, a better understanding of the chemical
abundances and evolution within dark molecular cores can be achieved.  As
rarer trace-molecules are routinely used to discern the distribution of
molecular material in star forming clouds, understanding the abundance and
chemistry of these molecules is vital to our understanding of star
formation.

We have obtained new radio observations of the rotational transitions of
$^{12}$C$^{18}$O, $^{12}$C$^{17}$O, $^{12}$C$^{32}$S, $^{12}$C$^{34}$S,
$^{13}$C$^{32}$S, and N$_2$H$^+$ towards the Northern Streamer in the dark
cloud (B168) associated with IC 5146. Using the near-infrared extinction data of
the same region published by \citet{lal99}, we present a detailed analysis
of the correlation between the molecular gas and the dust column density
in the IC 5146 dark cloud. In \S 2 of this paper, 
we describe the near-infrared data utilized in this study. 
In \S 3, we describe the
acquisition and reduction of the radio molecular line observations.
In \S 4 we examine the excitation and present molecular abundance profiles against
extinction.  In \S 5 we discuss the implications of these observations and
in \S 6, we detail our conclusions.

\section{Observations and Data Reduction}

\subsection{Near-Infrared Extinction}

The near-infrared data used in this paper are taken from \citet{lal99},
where a detailed description of the data acquisition, reduction, and
source extractions may be found.  The area surveyed is approximately
$27\arcmin \times 8\arcmin$, and the photometry is complete (at the
10$\sigma$ level) to a depth of $H \approx 17$ mag and $K \approx 16$ mag.
The infrared color excess for each star was determined via $E(H-K) =
(H-K)_{obs} - (H-K)_{intr}$, where $(H-K)_{intr} \equiv
\langle(H-K)\rangle_{control}=0.13\pm0.01$ magnitudes, the mean color
of field stars observed in a nearby, unextincted control field
\citep{lada94, lal99}. The
color excess for each star was converted to an extinction, using the
reddening law of \citet{rl85}: $A_K = 1.78E(H-K)$ \& $A_V = 15.9E(H-K)$.

The extinction derived in this manner is directly proportional to the {\em
true} dust column density along that line-of-sight, under the assumption
that the H$_2$-to-dust ratio remains constant.  The conversion of the
measured near-infrared color excess to the conventional visual extinction
may not truly represent the actual visual reddening of a background star,
because grain growth in cold clouds alters the extinction law at
$\lambda~\ll~1$~\micron.  However, at near-infrared wavelengths, the reddening law
is not known to vary significantly with grain growth 
\citep[e.g.,][]{math90}, and the derived near-infrared extinctions remain
proportional to the line-of-sight column density. 
Moreover, the conversion of the measured infrared extinction
to the conventional visual extinction maintains the  proportionality.  Use
of $A_K$ instead of $A_V$ would avoid this confusion, but to facilitate
comparison to other work \citep[e.g.,][]{lada94, crd98, alves98, kram99,
lal99}, we report visual extinction throughout this paper.

\begin{figure}
\plotone{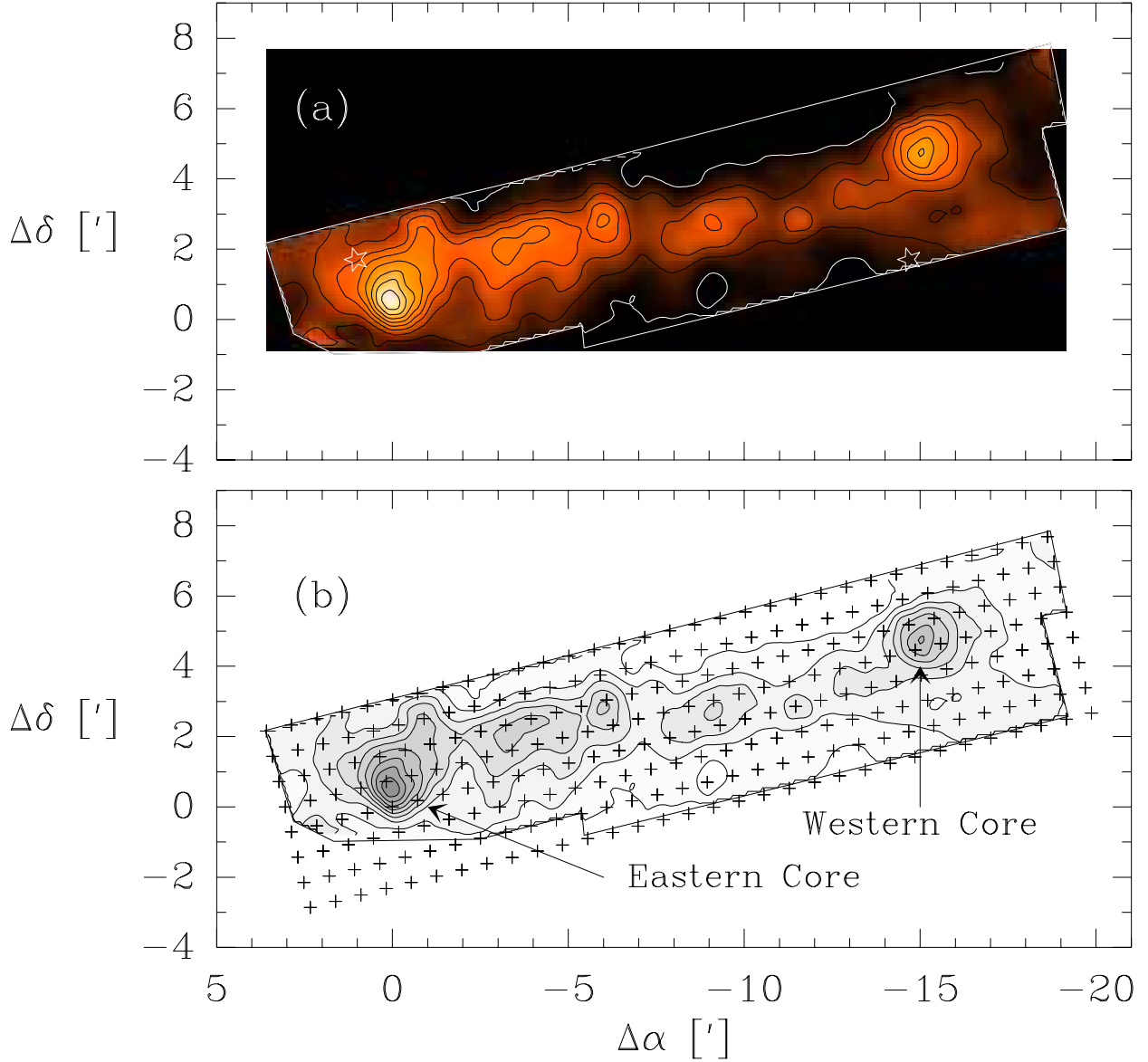}
\caption{
(a) Gaussian convolved map of visual extinction toward the northern streamer
of IC 5146.  Contours start at 4 mag of \av\ and increase in steps of 4 mag. 
The stars show the positions of IRAS point sources.
(b) Points delineating the radio grid overlayed on the region surveyed
in the infrared (as shown by the map of visual extinction).
The near infrared survey region is outlined in each.
The central reference position is listed in Table~1.
}
\label{fig_avgrid}  
\end{figure}

To enable a direct comparison between the near-infrared observations and
the radio molecular-line observations (obtained with the FCRAO 
telescope and described in the next section), the mean near-infrared
extinction along each line-of-sight associated with a molecular-line observation 
is calculated by convolving the individual infrared excess
measurements with a two-dimensional Gaussian filter.  The gaussian filter
has a FWHM of 50\arcsec\ to match the approximate beamsize of the FCRAO
telescope, and is truncated at $r=3\sigma$.  The uncertainty on the
mean extinction is estimated from the Gaussian weighted rms dispersion of
extinction measurements falling within each ``beam.''  
The gaussian convolved map of visual extinction is
shown in Figure~\ref{fig_avgrid}a.  This map is quite similar to the
larger extinction map of IC 5146 presented by \citet{lal99}. 

\subsection{Molecular-Line Observations}

Over two separate observing periods (1999 March 23--25 and 2000 Jan 31 --
Feb 01), the 14~m telescope of the Five College Radio Astronomy
Observatory (FCRAO) in New Salem, MA was used to observe a $25\arcmin
\times 8\arcmin$ region covering the Northern Streamer of the IC 5146
dark cloud.   The rotational transitions of the following molecules were
observed: $^{12}$C$^{18}$O (109.782182 GHz, $J =1\rightarrow$ 0; hereafter,
C$^{18}$O), the hyperfine triplet of $^{12}$C$^{17}$O (112.359277 GHz, $J
=1\rightarrow 0$; hereafter, C$^{17}$O -- note that the two reddest
components are blended and unresolved at our spectral resolution of 78
kHz; see below), $^{12}$C$^{32}$S (97.981011 GHz, $J =2\rightarrow$1;
hereafter, CS),  $^{12}$C$^{34}$S (96.412962 GHz, $J = 2\rightarrow$1;
hereafter, C$^{34}$S),  $^{13}$C$^{32}$S (92.4979 GHz $J = 2\rightarrow$1;
hereafter, $^{13}$CS), and  N$_2$H$^+$ (93.173777 GHz, $J =1\rightarrow 0$).  The
N$_2$H$^+$ transition consists of seven hyperfine components spanning 4.7
MHz; however, at our spectral resolution (78 kHz, see below) only the
three main groups of lines centered at rest frequencies of 93.1719 GHz (a
blend of three lines), 93.1738 GHz (a blend of  three lines), and 93.1763
(a single line) are resolved \citep[e.g.,][]{cmt95}.

The observations reported in this paper were obtained with the 
newly commissioned SEQUOIA 16-element focal plane array receiver.
The sixteen elements of the SEQUOIA array are aligned in a square
$4 \times 4$ pattern; each element is separated from its neighbor by
88\arcsec\ in both the X and Y directions.  A full beam sampled map
requires 4 pointings of the array on the sky and is referred to as a
``footprint''.   All maps were centered with respect to $\alpha(J2000) =$
21$^h$47$^m$27$^s$ and $\delta(J2000) =$ 47$^{\circ }31'00''$.

The FAAS autocorrelator spectrometer was utilized in the 512-channel, 40
MHz bandwidth mode, yielding a channel spacing of 78 kHz ($\Delta v \sim
0.21$ km s$^{-1}$ at 112 GHz and $\Delta v \sim 0.25$ km s$^{-1}$ at 93
GHz) The 1999 data were taken in position switching mode, while the
2000 data were taken in frequency switching mode.  For the frequency switched data,
the signal frequency was shifted by 4 MHz. Typical system temperatures
ranged from $140-300$ K. The integrations times for each molecule varied
(see Table 1), but typical $T_A^*$ rms noise temperatures per channel were
$\lesssim 0.02$ K. Internal calibration was done via a chopper wheel which
allows switching between the sky and an ambient temperature load. Focusing
and pointing were checked by observing periodically the SiO masers T Ceph
(1999) and R Cas (2000). Typical, rms pointing uncertainties were
5\arcsec.

The Northern Streamer was mapped at full beam sampling in C$^{18}$O, CS, and
N$_2$H$^+$.  The $25\arcmin \times 8\arcmin$ map was rotated by 14$^{\circ}$
North of East with respect to equatorial coordinates to follow the long
axis of the cloud and the near-infrared survey data
\citep[e.g.,][]{lada94, lal99}.  The full map required four ``footprints''
for a total of 256 grid points.  The streamer contains two dense cores
which are readily detectable in a 50$''$ beam and were mapped at full beam 
sampling in C$^{34}$S. The positions of the
grid points in these two individual ``footprints'' are no different than the
positions found in the the full map.  A single additional ``footprint''
(64 grid points) was observed towards the eastern core in C$^{18}$O. This
``footprint'' was offset from the regular map grid by $-0\farcm 369$
($\sim 0.5$ beamsize) in both right ascension and declination.   The
purpose of this ``footprint'' was to increase the sampling of the
C$^{18}$O in the eastern core. Additional single array placements (16
grid points) on the eastern core were observed in  C$^{18}$O, C$^{17}$O,
and $^{13}$CS.   These single array observations were used to investigate
the C$^{18}$O and CS optical depth of the core (see \S 3.1 and \S 3.2 below).

A summary of the map grid points is shown in Figure~\ref{fig_avgrid}, and details of the
observing parameters for each molecular observation can be found in Table
1. Several points on the southeastern edge of
the radio grid fall outside of the infrared survey region 
and these data points are not included in this study.   

For the remainder of the paper, the ``Eastern Core'' refers to the
eastern-most ``footprint'' (i.e., the eastern-most $8 \times 8$ grid),
the ``Western Core'' refers to
the western-most ``footprint'' (i.e., the western-most $8 \times 8$ grid),
and the ``Mid-Streamer'' refers to the two ``footprints'' ($16 \times 8$
grid) lying between the ``Eastern'' and ``Western'' Cores (see Fig. 1).

All data were reduced using SPA and CLASS spectral line reduction packages
in conjunction with custom-made analysis routines.  A second order
polynomial baseline was subtracted from the position-switch spectra; the 
frequency-switched spectra
were folded and a second-order polynomial baseline was subtracted.  Main
beam efficiencies of $\eta_{MB} \approx 0.50-0.55$ were adopted from the
SEQUOIA documentation.  Final $T_R$ rms noise temperatures were
generally $\lesssim 0.04$ K.  
All data presented in the paper are corrected using the main beam efficiency and are
on the $T_R$ scale.
%The spectral lines (including the individual
%components of the C$^{17}$O and N$_2$H+ hyperfine transitions) were all
%fit with single component Gaussian profiles.  Integrated intensity maps
%for C$^{18}$O, CS, C$^{34}$S, and N$_2$H+ are presented in Figures 2a-d.

\section{Results: Comparison of the Dust and Gas}

Figures  2a and b present maps of the integrated 
emission of \CeiO\ and \NtwoHp\ in 
IC 5146.  The map of \CeiO\ J $= 1 \rightarrow 0$ integrated intensity appears
broadly similar in distribution to the visual extinction map shown in
Figure~\ref{fig_avgrid}.  Both extinction and \CeiO\ trace column density 
enhancements, or cores, at the eastern and western edges of the mapped region,
although the \CeiO\ emission peaks are slightly offset from those of \av .
The integrated emission of \NtwoHp\ shows strong emission 
toward the eastern core, and only weak and somewhat clumpy distribution throughout 
the rest of the streamer.  Curiously, the \NtwoHp\ intensity appears to correlate
reasonably well with the regions of highest extinction.   This is more clearly demonstrated
in Figure~\ref{fig_wc} where the integrated intensity distribution of \NtwoHp\
and \CeiO\ are compared with visual extinction in the western core.  
Here the strongest \NtwoHp\ 
emission appears directly on the peak of extinction,  while the \CeiO\ emission 
maxima are offset by nearly 1$'$.  This is unlikely to be the result of differences 
in excitation requirements between these two transitions.  The upper state energy of
each transition is $\sim 5$ K and, while \NtwoHp\ has a large critical density 
($\sim 10^5$ \cc ) and preferentially samples only the highest densities, the
\CeiO\ emission should also trace the same density regimes.   Instead these 
morphological differences are likely due to differences in the chemistry as
will be discussed later in this paper. 

Figure~\ref{fig_csdist} presents the total integrated emission for the 
J $= 2 \rightarrow 1$ transitions of CS and \CtfS . 
The CS emission morphology is similar to that of \CeiO\ as 
CS strongly peaks on both cores and shows weak, but detectable, emission from 
the gas between the cores.  Because of its weaker intensity, the J $= 2
\rightarrow 1$ transition of \CtfS\ was only observed in the two cores. 
Within these smaller regions the
\CtfS\ emission is roughly similar to that of its more abundant isotope, CS.

In Figure~\ref{fig_spec} selected spectra at two different positions are presented.
The two chosen positions are ($\Delta \alpha = +0\farcm 2$, $\Delta \delta = +0\farcm 7$) and
($\Delta \alpha = -0\farcm 9$, $\Delta \delta = +2\farcm 5$) are both associated with the Eastern
core and sample extinctions that differ by a factor of two.  Here we see that the 
intensity of C$^{17}$O is significantly lower at the highest extinction (for \av\ = 36 mag the
detection is 16$\sigma$ on the integrated intensity, while at \av\ = 19 mag the 
detection is 4$\sigma$).  In contrast the emission of C$^{34}$S and \NtwoHp\ appears to
be stronger at the position with higher extinction.   
These differences, shown at only two positions, will be examined in
greater detail with higher sampling in the following sections.

\subsection{C$^{18}$O}

The direct comparison between the C$^{18}$O integrated intensity and
visual extinction at each point in the mapping grid
is presented in Figure~\ref{fig_c18oint}.   The behavior of these two
tracers is very similar to what was found previously for the IC 5146 region \citep{lada94,
kram99}.  The integrated intensity and the extinction appear 
well-correlated for $A_V
\lesssim 10-15$ mag.   For $A_V \gtrsim 10-15$ mag, the relationship
shows a high degree of scatter, and appears to flatten out.

A bivariate linear fit was performed for the data located at $A_V < 10$ mag, with the
following result:
\begin{equation}
I(C^{18}O) = \left(-0.4 \pm 0.1\ K\ km\ s^{-1}\right) + \left(0.20 \pm
0.01\ \frac{K\ km\ s^{-1}}{mag}\right)\cdot [A_V].
\end{equation}

\noindent The above fit is in excellent agreement with the fit presented by
\citet{lal99}, for a similar region but at lower angular resolution 
(102\arcsec) than the observations presented here. 
The fit is also in
agreement with the C$^{18}$O-$A_V$ relationship found for the smaller
region (and finer beamsize -- 30\arcsec) studied by \citep{kram99}.  
However, if the entire extinction range is used in the fit, the resulting slope is
15\% shallower, consistent with earlier findings for both IC5146
\citep{lada94} and also in L 977 \citep{all99}.

The two most likely effects which could explain the change in slope 
at high extinction
are: (1) high C$^{18}$O optical depth for $A_V \gtrsim 10$ mag, and (2)
depletion of CO onto the surface of dust grains.   A third possibility is
a decrease of the CO excitation temperature to $T_{ex} < 5$ K for $A_V
\gtrsim 10$ mag.  However, \citet{all99}, for a similar result in L~977,
make a convincing argument that this is not a likely possibility. They argue that
heating mechanisms (cosmic ray rates, ambipolar diffusion, and natural
radioactivity) are sufficient to prevent the gas in cloud cores to cool
below 5 K.  Although we cannot completely rule out this possibility, we
regard the cooling of the gas to such extreme temperature as unlikely, and
proceed with a discussion of the optical depth and depletion.

The paired C$^{18}$O--C$^{17}$O single pointing observations provide
information regarding the optical depth of the C$^{18}$O in the densest
portions of the eastern core (see Table 1).  C$^{18}$O and
C$^{17}$O should be collisionally excited under the same physical
conditions deep inside cold cores where the molecules are well-shielded
from external ultraviolet radiation \citep{lfd98, vdb88}.  Additionally,
the ratio [$^{18}$O]/[$^{17}$O]$=3.65 \pm 0.15$ appears to be relatively
constant throughout the interstellar medium \citep{penz81}.  Thus, the
only difference in the emission between C$^{18}$O and C$^{17}$O should
arise from the oxygen relative abundance; i.e., completely optically thin
emission of C$^{18}$O and C$^{17}$O should have an integrated intensity
ratio of 3.65.

A comparison of the integrated intensities of C$^{17}$O vs C$^{18}$O is
shown in Figure~\ref{fig_c17o} (top).  
The solid line represents the expected ratio of
3.65 for optically thin emission and does {\em not} represent a best fit
to the data.  The dashed lines represent the $1\sigma = \pm0.15$
uncertainty of the [$^{18}$O]/[$^{17}$O] ratio.  The excellent agreement
of the data distribution with the predicted ratio indicates that the
C$^{18}$O emission is indeed optically thin. 

To test directly for optical depth effects as a function of total column
density, the C$^{18}$O/C$^{17}$O ratio has been plotted as a function of
extinction (bottom, Fig.~\ref{fig_c17o}).  The extinction provides an independent
assessment of the total column density along each line of sight.  If the
C$^{18}$O was optically thick at high column density (high extinction), the
C$^{18}$O/C$^{17}$O ratio should systematically decrease as a function of
increasing extinction.  The ratio is remarkably constant over a range of
40 magnitudes of visual extinction.  Figure~\ref{fig_c17o} implies that (1) the
[$^{18}$O]/[$^{17}$O] ratio in IC 5146 is very near the previously
determined interstellar value of $3.65 \pm 0.15$, and (2) the C$^{18}$O
emission is {\em not} highly optically thick.  However, due to the errors on the
ratios we cannot rule out the C$^{18}$O emission being thin, but with moderate
optical depth ($\tau \lesssim$ 0.5 ).  

The ratio of the three hyperfine components of C$^{17}$O can be used to
determine whether or not C$^{17}$O is thin. 
The relative intensities of the
three components (central:blue:red) are 4:3:2 for optically thin emission
\citep{lfd98}.  At the
spectral resolution of our data (78 kHz), the central and red components
are blended; thus, the measured relative intensity of the two components
(central-red blend:blue) should be 2:1, for optically thin emission 
and indeed the measured median
relative intensity is $2 \pm 0.5$.  If this is the case then the difference
in the two  C$^{17}$O spectra presented in Figure~\ref{fig_spec}, with the highest extinction
having lower emission, is suggestive of a drop in abundance.

\subsection{CS}

The depletion of sulfur bearing molecules (e.g., CS and SO) is predicted to be
a sharp function of the density, and these species should be robust tracers
of gas phase molecular depletion on to the surface of dust grains
in cloud cores \citep{bl97}.  As there
was already evidence to suggest that CO may be depleted in the cores of IC
5146 (\S 3.1), CS should also show evidence for gas-phase
depletion.

To explore this hypothesis, the C$^{34}$S optical depth was computed for
both the eastern and western cores at positions where we have observations
of both CS and \CtfS , using the following relationship:
\begin{equation}
\frac{\rm T(C^{34}S)}{\rm T(CS)} = \frac{1-\exp{[-\tau]}} {1-\exp{[-a\tau]}},
\end{equation}
where $\tau$ is the C$^{34}$S optical depth and $a$ is the abundance ratio
of $[^{32}\rm{S}]/[^{34}\rm{S}] = 14$ \citep{pratap97}.  Using the ratio of the
integrated intensities, we iteratively solved  Equation 2 for $\tau$, until
the calculated intensity ratio matched the measured intensity ratio to
within $\sim$5\%.  In Figure~\ref{fig_cstau}, the C$^{34}$S optical depth is shown
as a function of visual extinction $A_V$.
The uncertainties shown in Fig.~\ref{fig_cstau} are a result of the
uncertainties in the integrated
intensity measurements; the uncertainties associated with the
convergence of the numerical calculations are significantly smaller.
The derived opacities clearly indicate that the emission from C$^{34}$S is
optically thin in both cores for all A$_V$.  However, the emission from CS ($\tau_{\rm CS}
= \tau_{\rm C^{34}S} * 14$) must be optically thick throughout the cloud.
Surprisingly, despite being thin, the C$^{34}$S optical depth 
is nearly constant for $A_V \gtrsim 10$ mag.   

As a check of the derived optical depths, the optical depth of $^{13}$CS
was calculated in a similar manner to the C$^{34}$S optical depth.  Only
the position ($-0\farcm 54, 0\farcm 89$) was sufficiently strong to permit
such a calculation ($\int T(^{13}{\rm CS}) dv = 0.089\pm0.02$ K km
s$^{-1}$).  An optical depth of $\tau = 0.033 \pm 0.008$ was found,
implying that the C$^{34}$S optical depth (assuming an abundance ratio of
$[\rm{C}^{34}S]/[^{13}\rm{CS}] = 4.29$) is
$\tau = 0.14 \pm 0.03$ (see Fig. 5).  The C$^{34}$S optical depth 
measured using the T(\CtfS )/T(CS) ratio is $\tau = 0.14 \pm 0.04$, 
at the same position, confirming the small opacities of C$^{34}$S 
in this cloud.

Examining the dependence of \CtfS\ opacity with extinction
in Figure~\ref{fig_cstau} we find that while the extinction rises 
by a factor of 2.5 ($15-40$ mag), the C$^{34}$S
optical depth remains nearly constant at $\tau \approx 0.15-0.25$.  Indeed, there
is even a small hint that the optical depth of the C$^{34}$S in the
eastern core is decreasing as a function of increasing extinction.
The eastern core also appears to have an opacity that is, on average,
slightly higher than the western core.
As optical depth is a direct tracer of column density in either the 
upper or lower state, this difference suggests that the abundance
of CS could be different between the two cores. 
Moreover, the constant opacity with extinction for both cores
would then be indicative of 
a \CtfS (and CS) abundance decrease with increasing extinction. 
However, CS and \CtfS\ have high dipole moments and their emission 
is sensitive to gas density, as opposed to
\CeiO\ emission which is fairly insensitive to the density (provided
\nhtwo\ $> 1000$ \cc ).  Thus, for CS and its isotopic
variants, changes in the excitation conditions can play an important,
and with regards to depletion analyses, potentially confusing role. 
These effects are examined in \S 4.

\subsection{N$_2$H+}

Unlike CS, or even CO, \NtwoHp\ is predicted by evolutionary
chemical models to have a low depletion rate, due to the 
relatively low binding energy of its precursor molecule, \Ntwo\
\citep{bl97}.  
Thus, we might expect that \NtwoHp\ would exhibit a different
behavior as a function of extinction when compared to CS or CO.

In Figure~\ref{fig_n2hpint}, the 
the N$_2$H+ integrated intensity, summed over all hyperfine transitions,
is compared to the visual extinction.   This figure clearly shows 
quite a different relationship between \NtwoHp\ integrated emission and 
visual extinction when compared to
the other species included in our study.
For $A_V
\lesssim 15$ mag, the N$_2$H+ integrated intensity 
vs $A_V$ is nearly flat and follows the
following relationship:
\begin{equation}
I(N_2H+) = \left(-0.03 \pm 0.04\ K\ km\ s^{-1}\right) + \left(0.017 \pm
0.006\ \frac{K\ km\ s^{-1}}{mag}\right)\cdot [A_V]\ (A_V < 15\ {\rm mag}).
\end{equation}
Note that the slope of the linear fit is non-zero at the 3$\sigma$ level,
indicating that the N$_2$H+ intensity gradually increases with extinction.
At $A_V \gtrsim 15$ mag, the intensity displays a sharp increase and the 
N$_2$H+$-A_V$ relationship steepens by nearly a factor of 15:
\begin{equation}
I(N_2H+) = \left(-4.3 \pm 0.6\ K\ km\ s^{-1}\right) + \left(0.26 \pm
0.03\ \frac{K\ km\ s^{-1}}{mag}\right)\cdot [A_V]\ (A_V > 15\ {\rm mag}).
\end{equation}
Curiously, the increase in \NtwoHp\ intensity occurs at a similar
visual extinction to where the CO-\av\ relationship begins to show
evidence of potential depletion.   
The derivation of \NtwoHp\ abundances and the physical cause for the
sharp increase at \av\ $\gtrsim 15$ will be discussed in the following 
section.

\section{Analysis: Molecular Excitation and Abundances}

In the previous section we searched for correlations between integrated
emission and visual extinction for each of the surveyed molecules.   
Here we examine whether any departures from the correlation 
are the result of excitation and/or abundance gradients.
In this section we concentrate on trends in relative or normalized
abundances (i.e. abundances normalized using the abundance derived at the lowest extinction with
reliable data points).  In this fashion, 
in the determination of molecular abundance for a given species, 
we eliminate common uncertainties (such as collision rates).   

To calculate total column densities from the molecular data we
account for the radiation transfer through the use of the
large-velocity gradient approximation.   To reduce the effects of opacity
we use here only the C$^{17}$O, \CtfS , and \NtwoHp\ data.
Since the emission of C$^{17}$O and \CtfS\ (\S 3.1 and 3.2) is optically
thin, and, based upon ratios of hyperfine components, much
of the \NtwoHp\ emission is also thin, this 
reduces the LVG approximation to simply solving the equations of statistical equilibrium. 
The results are therefore less dependent on the details of the radiative transfer solution.
To improve the statistics of our results we average abundances in bins of 
5 magnitudes starting with \av\ = 5 mag.  Below this value there are little or no significant
($> 3\sigma$) data points (for \NtwoHp , C$^{17}$O, and \CtfS ).

In this study we observe only a single transition of a given molecular species.
However, in each case we have information on the optical depth, using
CS for \CtfS , and the hyperfine ratios for \CsvO\ and \NtwoHp .   
This provides an
additional and limiting constraint on the column density determination.
With the observed intensity and opacity from a single transition the total column density can
be derived with knowledge of the collision rates, density, temperature, and line width. 
The collision rates are available from the literature
and the line width or velocity dispersion is an observed
quantity.  However, to derive a map of the total column density from
the molecular line maps requires
a priori knowledge of the density and temperature structure of the
IC 5146 cloud.  

For the density structure we use the analysis in (Lada, Alves, \& Lada 1999; hereafter LAL99)
who show that the extinction (or column density) gradient in the same
region of IC 5146 is nicely 
reproduced by a cylindrical geometry (for the entire streamer)
with $\rho \propto r^{-2}$.  The procedure LAL99 use to determine the density structure
is identical to the commonly used method of deriving the volume density structure 
from mm and sub-mm dust continuum emission (see Shirley et al 2000 and references
therein).  To be consistent with the results in LAL99, which show
a large systematic increase in volume density from low to high extinction, we 
adopt the following density profile (\nhtwo\ $=$ \nhtwo (r = r$_0$) $\times$ (r$_0$/r)$^{2}$
$= 10^5$(0.047 pc/r)$^2$ \cc ).  The density profile is shown in Figure~\ref{fig_denpro} and is
in reasonable agreement with the radial profile of total column density (Figure 8 in LAL99).
In our calculations
we use average densities in bins of 5 mag (shown as horizontal hash marks in
Figure~\ref{fig_denpro}).  For example, if a given data point has an extinction between 5 and 10
mag then we use the average density of \nhtwo\ $= 1.4 \times 10^{4}$ \cc\ within that bin.

There exists some information on the temperature structure from
observations of CO and NH$_3$.  CO observations generally show gas temperatures ranging from
10 to 13 K \citep{dobashi92}.  However, these estimates are likely reflective of the 
temperature at the cloud surface and not the denser interior \citep{bgsu94}.  
In the eastern core NH$_3$ observations provide some information on the dense gas
temperature with T = 13 K \citep{jma99}.  We also have a single detection
of the J = 6 $\rightarrow$ 5 (K = 0 and 1) transition of CH$_3$C$_2$H, which can be used to
estimate the temperature \citep{bgsu94}.   Since this detection is at a single position
($\Delta \alpha = -0\farcm 5$, $\Delta \delta = +0\farcm 9$)
we have not shown the spectra; however, the K = 0 integrated intensity 
is $\int T \delta v =$ 0.11$\pm 0.02$ K km s$^{-1}$ and K = 1 is 
$\int T \delta v =$ 0.13$\pm 0.02$ K km s$^{-1}$.  
Using the method described in \citet{pratap97} the ratio
of these two is consistent, within the errors, with a gas temperature $< 13$ K.  
Given this information, it is unlikely that there are strong temperature gradients
throughout most of the IC5146 molecular streamer 
with the probable range between (perhaps) 5 and the CO temperature
of 13 K.  In the following we present abundances calculated with a constant temperature
of 10 K and discuss the results of additional analysis assuming constant temperatures
of 5 and 13 K (and combinations thereof).  

\subsection{C$^{17}$O Abundance}

Because of concerns that moderate opacities in \CeiO\ emission could potentially
mask small abundance changes in order to search for relative abundance differences
we predominantly use the more limited \CsvO\ observations.
For the \CsvO\ column density derivation, we use the collision rates of CO with para-H$_2$
\citep{f88}.  We adopt the density determined as a function of extinction 
as described above and the temperature is assumed to be constant at 10 K for each position
(changes in the temperature will be discussed below).
The line width is determined via gaussian fits to
each spectrum. 
Because the \CsvO\ J $= 1 \rightarrow 0$ transition has hyperfine structure we use 
this additional information in our derivation of column density.  We therefore
simultaneously fitted the two resolved hyperfine components (as discussed in \S 3.1 the
F $= 7/2 \rightarrow 5/2$ and F = $= 3/2 \rightarrow 5/2$ are blended), assuming that the
hyperfine levels are populated according to LTE.  The two constraints on the
on the $\chi^2$ search for the best fit column density are: (1) the intensity of the lines
and (2) the relative intensities of the hyperfine ratios which limits the opacity of the
solution.  In general the reduced $\chi^2$ of most solutions are $< 1$ and
in all cases the \CsvO\ emission is found to be optically thin. 
For each position we derive column densities if the observed integrated intensity is $> 3\sigma$.   
The abundance relative to H$_2$ is derived using N$_{H_2} = 10^{21}$\av .

Table~3 presents the weighted average abundance of C$^{17}$O 
(within 5 mag bins of extinction) and in Figure~\ref{fig_abun} 
the relative abundance of \CsvO\ is given as a function
of visual extinction.   In this figure we normalize the abundance to the lowest
extinction bin with a statistically significant abundance determination (between 5 and 10 mag). 
Excluding the point at \av\ = 27.5 mag with large errors,
we see that the relative \CsvO\ abundance shows a steady and significant decrease, by
a factor of 3, from the lowest to highest visual extinction.
This result is found assuming a constant temperature of 10 K.  To investigate
whether a systematic temperature gradient can change the result we repreated the same
procedure assuming constant temperatures of 5 and 13 K.  To mimic temperature gradients
we examined combinations of these results.  In the absence of a significant population
of luminous embedded stars (the case here) the most likely systematic temperature gradient
is the temperature decreasing with increasing extinction.  In this case we find that
the \CsvO\ abundance at high \av\ (with T = 5 K) does increase due to higher 
opacity in the J = 1 state, allowing for greater column density, but by only 15\%.   
At the cloud edges (provided T = 13 K) the abundance also shows a slight increase,
primarily due to the lower population in the J = 1 state.  If the temperature were even
greater then the abundance continues to rise.  In all, this examination
still provides a factor of 3 decrease in the \CsvO\ abundance from \av\ = 5 to 40 mag.  
It is worth emphasizing that even if CO and its isotopic variants are completely depleted
in the dense core center the molecular observations will detect emission from the relatively
undepleted gas residing in the low density regions along the line of sight through the core.
Thus, if CO shows evidence for depletions at 10 mag in a cloud with a total extinction (from
Fig~5 for example) of 40 mag, then the measured abundance will drop by at most a factor of four. 

A similar analysis was performed using the entire \CeiO\ data set and we find that the 
relative abundance is essentially constant until the highest extinction bin, whereupon the
abundance decreases by a factor of 1.5.  This confirms our suspicion that the \CeiO\ 
emission is likely to have moderate optical depth ($\tau \sim 0.5$) which masks
the larger abundance difference (factor of 3) seen in \CsvO . 
We note that one core in our data set ($\Delta \alpha = -9\farcm 2$, $\Delta \delta = 2\farcm 9$)
has been examined with the IRAM
30m antenna at higher ($\sim 30''$) resolution by \citet{kram99}.
They find a clear decrease in the CO abundance near 12 mag.    
Our results are not as dramatic as found in the higher resolution study, which  
allowed for more data points to be placed at higher extinctions.  
For instance, the \citet{kram99} core
shows a peak extinction in a 30$''$ beam of $\sim 28$ mag,
at a resolution of 50$''$ this reduces to a peak extinction of 16 mag.
Thus, our sampling is lower in the exact regions where gas-phase 
depletions are most likely.  
%Moreover, there are several points near 
%the centers of dense cores that, due to a lack of stars within the 
%central portions of the gaussian NIR beam, we are able to derive only 
%lower limits to their extinction.
 
\subsection{\CtfS\ (and CS) Abundance}

To derive the total column density of \CtfS\ we use a similar 
method described for \CsvO\ in the previous section.  That is, with
a density and temperature assumed (and constant) for each grid
position we perform a $\chi^2$ search for the best fit total column
density.  For collision rates we use the rates given by
\citet{gc78}, and the line width is
determined via gaussian fits to the spectrum.
\CtfS\ does not have hyperfine structure, but we have additional information 
in that the \CtfS\ opacity has been determined for all positions 
with $> 3\sigma$ integrated intensities (\S 3.2).  Thus,
for each data point we again 
two constraints for the radiative transfer model: (1) the integrated intensity
and (2) the opacity of the J $= 2 \rightarrow 1$ transition.   
The majority of column density solutions have reduced $\chi^2 < 1$.

The \CtfS\ integrated intensity as a function of visual extinction is 
presented in Figure~\ref{fig_c34sint}.  This Figure is notable in that
the \CtfS\ integrated intensity appears to be reasonably well correlated
with \av .  This result is somewhat surprising given the near constant
opacity as a function of extinction shown in Figure~\ref{fig_cstau}. 
The near uniform opacity would suggest that the \CtfS\ (and therefore CS)
abundance is sharply decreasing with extinction.  However the dependence 
in Figure~\ref{fig_c34sint} is at odds with that conclusion,   
as it requires a constant abundance.
This difference, an integrated intensity correlated with \av , combined with a
constant optical depth, can be reproduced, 
provided that there was a change in excitation
from low to high extinctions.  
Indeed we can expect such a increase in the \CtfS\ excitation temperature 
proceeding from lower to higher \av , as a direct result of the known density gradient.
Even a small change in the excitation temperature could replicate the observed
dependence.  

The results of the \CtfS\ excitation analysis are presented in Table~3 in the form
of abundances and in Figure~\ref{fig_abun} as normalized abundance 
plotted against visual extinction.   Examining the data points we find 
a steady statistically significant abundance ($> 3\sigma$)
decrease by an overall factor of 3 from low to high extinction. 
Thus, by accounting for the rise in density as required by the NIR observations, and
the observed constant opacity, along with an integrated intensity that increases with 
extinction, we find that the CS abundance decreases with \av .  
If we assume that the most likely temperature gradient is with warm gas at core edges 
(T = 13 K)
and cold gas (T = 5 K) in the center then the abundance still systematically declines 
by a factor of 2.  In all, these results suggest that CS, like CO, shows an abundance 
decrease in the dense gas likely due to depletion onto grains. 

\subsection{\NtwoHp\ Abundance}

To derive the abundance of \NtwoHp\ we use the total intensity of the
lines along with the hyperfine ratios as described  earlier for C$^{17}$O. 
We use the collisional
rates for HCO$^+$ excited by para-H$_2$ \citep{f99}, which  
has comparable collisional rates to \NtwoHp\ \citep{m84}. 
The results of the excitation calculations are given in Table~3 and shown
in Figure~\ref{fig_abun}.

First, there are only 3 \NtwoHp\ detections 
between 0 and 5 mag of extinction, and of these, two have
\av\ $>$ 4.5 mag (these data are not provided due to high errors).  
Although \CtfS\ and \CsvO\ also have little emission
in this regime, the more abundant isotopes, \CeiO\ and CS, both are detected
at low extinction. 
Thus, unlike for other species, there appears to be little \NtwoHp\ below
\av\ $\lesssim$ 4 mag.  
Beyond the lowest extinctions, for T = 10 K the \NtwoHp\  normalized abundance
in Figure~\ref{fig_abun} shows a high abundance for low \av\ ($\sim 7.5$ mag)
then decreases by a factor of 3 for \av\ = 17.5 mag, whereupon the abundance
increases by a similar factor at high \av .  In our solutions with different
temperatures (T = 5, 10, and 13 K) we find that the initial abundance decrease
for \av\ $< 17.5$ mag is lessened, provided that the cloud is warmer in these regions.
If the gas is cooler in the core center then we find an even greater increase
in relative abundance for \av\ $> 17.5$ mag. 

In summary we find, at least, two regimes for the \NtwoHp\
abundance:  (1) there is a threshold of extinction, $A_V^{th} \lesssim 4$
mag below which we find little or no \NtwoHp\
in the gas-phase and (2) for higher extinctions the abundance shows a complicated
structure that initially declines in value until \av\ = 15 mag whereupon
it the relative abundance increases with extinction.  Consideration of potential temperature
changes reduces the initial decline, but would increase the rise in relative abundance
towards to the core center.  However, there is 
no evidence for a systematic decrease in the \NtwoHp\ abundance at large \av .
In the following section we examine the physical and chemical
causes that give rise to this dependence.
 
\section{Discussion}

\subsection{Differential Molecular Depletions in IC 5146}

In the preceding section we examined the excitation of the three
molecules included in our survey.  The results of this analysis is
that we find good evidence for the presence of differential molecular
depletions in IC 5146.  That is, \CeiO\ and CS (and \CtfS ) are 
depleting from the gas phase at high extinctions and \NtwoHp\ is not (at least at a
resolution of 50$''$).    Similar evidence of differential
depletions are found in the literature, for a summary see \citet{bergin00}.  
The depletion of CCS relative to both ammonia and \NtwoHp\ (with the 
latter two tracers typically coincident with the dust continuum emission
peak) found in isolated low-mass cores is quite similar to the 
results found in our work \citep{klv96, olwh99}.
Recently, \citet{caselli99} found evidence of
CO depletion in the starless L1544 cloud core in a region traced by
both \NtwoHp\ and dust continuum.
The detection of gas phase molecular depletions should not be considered as unexpected,
given that molecular ice features have been observed
along numerous lines of sight in the ISM \citep{whittet93, ttgb91}.

In general, previous searches for
chemical differences related to molecular depletion relied on 
finding morphological dissimilarities between the emission of
two or more species with each other or with dust continuum
(similar to that shown in Figure~\ref{fig_wc}) and 
then performing an excitation analysis to prove or disprove that
such differences are the result of chemical abundance variations.
Our study has found comparable results, but has placed the evidence
for molecular abundance changes, which are attributed to
depletion, on firmer statistical grounds and removes any ambiguity in terms
of the total hydrogen column density, such as those associated with
using dust continuum emission \citep{kram98}.\footnote{Of course, another
method to search for molecular depletions is to directly observe
molecules in the solid state.  This method works quite well for
the dominant molecules on grain surfaces (such as H$_2$O or CO).
However, it fails for lesser abundant species such as CS because, even 
if they depleted entirely from the gas-phase their abundances are too
low to produce observable absorption features.}
 
\subsection{Comparison with Chemical Theory}

This depletion pattern with sulfur bearing molecules
(e.g. CS, CCS), along with CO, showing depletions onto grains and 
\NtwoHp\  remaining in the gas-phase is in good agreement with the
theoretical models presented by \citet{bl97}.  In these models,
the gas-phase chemistry includes the effects of
molecules both depleting onto and evaporatoring from grain surfaces.
\NtwoHp\ remains in the gas-phase longer at higher
densities than other molecules  because the high 
volatility of its pre-cursor molecule,
\Ntwo , allows for various desorption mechanisms (such as thermal
evaporation or cosmic-ray spot heating of grain surfaces) to keep
a significant \Ntwo\ abundance, and therefore \NtwoHp ,
in the gas phase at high densities.   
%By comparing
%our abundance structure as function of extinction (Figure~\ref{fig_abun})
%to these same chemical models
%we have found that CO depletion is also an important factor in allowing
%\NtwoHp\ to preferentially trace only the high density (and high \av ) gas. 

To compare our results with the gas-grain chemical model of \citet{bl97},
we use the density profile given in Figure~\ref{fig_denpro}.  This profile
matches the one derived for IC 5146 from the NIR extinction measurements 
(LAL99).  We adopt the notation $\tau_V$ with regards to cloud depth
because the model is computed using cloud radius, thus \av\ $= 2\tau_V$. 
With extinction and density (we also assume a constant dust and gas temperature
of 10 K and a UV enhancement factor of $G_{0} = 2.2$; Kramer et al 1999)  
predetermined as a function of radius,
we use the chemical model of \citet{bl97} to predict abundances as a function of 
both time and visual extinction and then directly compare these to the observational
results.  This model is slightly different from \citet{bl97} in that we
do not allow for the density to evolve with time, but rather
hold the density constant for each radius (and therefore extinction).
However, the density does increase with increasing extinction, properly
matching the observed structure.

Figure~\ref{fig_chemmod} presents two different chemical models.
The top two panels show a pure gas-phase
chemical model sampled at two different times ($10^5$ and 10$^7$ yr).
This model includes the effects of grain absorbing 
UV radiation (thereby allowing for molecular formation) but does not
include any gas-grain interactions.  
We note that to match the models to observations $\tau_V$ must be
multiplied by a factor of two to account for both sides of the cloud
(i.e. the model was done using radius and not diameter).
In Figure~\ref{fig_chemmod}, we see that the abundances of most
species rise as the UV radiation field is attenuated
with increasing extinction and then remain constant.  This behavior
exists despite the nearly two order of magnitude rise in density from edge
to center.  However, for \NtwoHp\ there is a slow steady decrease in abundance
towards greater extinction.  This behavior is readily understood by examining
the primary formation and destruction processes for \NtwoHp .
Using the pathways outlined in Appendix A, we see
that the formation rate of \NtwoHp\ (via cosmic ray ionization) should 
be essentially constant with density, with the assumption that cores are
threaded by cosmic rays.  However,
due to the higher probability for collisions with higher density,
the destruction rate via dissociative recombination and reactions
with CO is raised. Thus, for pure gas-phase models the prediction 
is that the abundance of \NtwoHp\ decreases with extinction.

The second model shown in the bottom two panels of Figure~\ref{fig_chemmod}
is a gas-grain chemical model where molecules are allowed to collide
and stick to the surfaces of dust grains.  The primary desorption mechanism
is cosmic-ray spot heating.  For
the binding energy of molecules to the grain surfaces we assume that
the grains are coated by a layer of CO molecules.  For more details
of the model see \citet{bl97}. 
Comparing the pure gas-phase model with the gas-grain model at t = 10$^{5}$ yr 
the abundance profiles are similar.  At this
early time depletion has yet to play a role.  For later times, we
see demonstrative effects due to grain depletion: (1) 
for $\tau_V > 6$ mag (\av\ $> 12$ mag) the abundance of CS is
dramatically reduced (2) the abundance
of CO shows a small, factor of $\sim$3, decrease between $\tau_V$ $= 0$ to
16 mag and (3) the abundance of \NtwoHp\ is constant with \av .  For \NtwoHp\ 
this behavior is in direct contrast to that seen in
the pure gas-phase case.   This indicates
that the \NtwoHp\ destruction rate must be progressively lower with 
higher extinction and density in the
gas-grain model than for pure gas-phase chemistry.   From Appendix A the major
destroyers of \NtwoHp\ are CO and electrons.  In Figure~\ref{fig_chemmod}
there is increasing CO depletion with higher density and extinction 
which results in a progressively lower \NtwoHp\ destruction rate.
Thus, the observed uniform abundance of \NtwoHp\ at high extinctions is
also a direct consequence of CO depletion and is not solely the result of
the high volatility of the \Ntwo\ molecule.
It is worth noting that if more CO molecules freeze onto grains, as would occur when
time progresses, then \NtwoHp\ abundance rises accordingly. 

Finally, we note that at low \av ,
for CS and CO (x(\CeiO ) $\times$ 500), 
the models can be directly compared to the observed abundances and
predicted values are in reasonable agreement with observations
(compare Figures~\ref{fig_abun} and~\ref{fig_chemmod}). 
For high \av , the models, as presented here as abundance profiles
with extinction, cannot be directly compared to the observations.
This is because the observations are averages over the entire line of 
sight, including
both depletion zones and undepleted gas.   For a direct comparison,
the chemical model would have to be
averaged over the line of sight, stepping through \av , 
in similar fashion to the observations.
However, this is not the case for undepleted species (e.g. \NtwoHp ), which
can be directly compared to observations.
If the relative abundance profile in Figure~\ref{fig_abun} is correct then
the observations would suggest a combination of undepleted models (to account
for the low \av\ decline in abundance) and the depleted models (accounting for
the rise in abundance at high extinction) would be required. 
In each case the models are in reasonable agreement with the 
overall value of abundance.

\subsection{Testing Interstellar Photodissociation Rates}

In \S 4.3 we found two regimes for the \NtwoHp\ abundance:
(1) there is a threshold extinction, $A_V^{th} \lesssim 4$ mag, below
which there is little or no \NtwoHp , and
(2) for higher extinctions the abundance is roughly constant.  In
the previous section we discussed how the structure in
latter regime can be reproduced.  However, in Figure~\ref{fig_chemmod} the
dependence found in the former regime is also addressed.   
In the chemical model 
the lack of \NtwoHp\ for \av\ $\lesssim 4$ mag
is the result of the photodissociation of its parent molecule (\Ntwo ),
combined with a higher electron abundance at cloud edges due to
photoionization. 
In Figure~\ref{fig_chemmod}, \NtwoHp\ shows a sharp rise in abundance for 
$\tau_V \sim 3$ mag, corresponding to \av\ $\sim 6$ mag.  This is 
quite close to the observed threshold.   The 
\Ntwo\ photorate is taken from \citet{vd88}, where the depth dependence of
the photodissociation rate
depends on its absorption cross section\footnote{For \Ntwo\ predissociation occurs
mainly between 912--1000 \AA\ \citep{carter72, rtt81}}, the assumed radiation
field, the continuum attenuation, and on the albedo and 
scattering phase function of the grains \citep{vd88}.  Despite the
complications involved in this computation, given the match
between the observed and predicted threshold, the depth dependence of the
rate and the ionization structure, appears to be reasonably taken into account. 

We note that the chemical model in Figure~\ref{fig_chemmod} 
does not properly account
for the self-shielding of CO molecules, as \CeiO\ is emission is observed
for \av\ $< 4$ mag (Figure~\ref{fig_c18oint}).  However, the effect of
increased CO shielding in the model would be to lower the electron abundance at the
cloud edge and increase the importance of \Ntwo\ photodestruction 
as the method for lowering the \NtwoHp\ abundance.   Our assertion 
that \Ntwo\ photodissociation is contributing to the observed \NtwoHp\ abundance
threshold would then be strengthened.

\citet{vd88} present a different photorate computed using a different 
grain model than used in our model, one with the grains more forward
scattering (their grain model 3).  If we use the depth dependent photorates from
that model (for all steps in \Ntwo\ and \NtwoHp\ production) we find that
the \Ntwo\ photodissociation rate is larger deeper into the cloud and
the \NtwoHp\ threshold shifts to beyond \av\ = 10 mag. This is in  
disagreement with the observations and points to an important use of the 
technique outlined in this paper.  
In the future, by observing and modeling clouds with simple and
well-characterized structure, the depth
dependent photorates (along with the gas-grain interaction) can be tested.
Indeed, most of the depth dependent photorates for many molecules are
highly uncertain \citep{vd88}.  In some cases this is due to a lack of viable
cross-sections, but also because the detailed rate calculations could
not be observationally verified. Although, there certainly are complications in 
application; this technique offers a viable opportunity to
provide observational tests of these rates, which can help in constraining
not only the depth dependence of the photorates but also the grain physics.

\section{Summary}

We have examined the correlation of the emission of three molecules:
\CeiO , CS, and \NtwoHp\ with visual extinction in the northern streamer
of the IC 5146 cloud.  We use the molecular data, along with a model
of the excitation, to determine molecular abundances and examine the
abundance profile with \av .  The principle results are:

(1) We find good evidence for a reduction in the abundance of CO and
CS (using \CtfS ) for \av\ $\gtrsim 10 - 15$ mag.   This reduction is
attributed to the depletion of these molecules onto the surfaces of cold
dust grains in the densest portions of the IC 5146 cloud.

(2) For \NtwoHp , we find two separate regimes for its abundances: (a)
for low extinction there is a threshold extinction, $A_V^{th} \lesssim 4$ mag,
below which little \NtwoHp\ is found and (b) for \av\ $\gtrsim 4$ mag the
\NtwoHp\ abundance is either constant or declining until \av\ = 15 mag whereupon
the abundance rises as a function of extinction in the visual.

(3) We find that density gradients in cloud cores can have a 
profound effect on the derivations of the spatial distributions
of total column densities and molecular abundances for species with
high dipole moments.  Chemical interactions, such as reaction rates
and depletion timescales, are also strongly density dependent.
Thus the interpretation and determination of accurate abundances in dense 
cloud cores requires both knowledge of the density and temperature
structure in the cloudy material and a proper accounting for such 
structure in abundance determinations and chemical models.

(4) The observed patterns of differential depletion shows CS and CO exhibiting
molecular depletions while \NtwoHp\ remains in the gas phase. This is in good
agreement with the predictions of chemical theory by \cite{bl97}.  

(5) By combining the observationally derived density profile for IC 5146 from
\cite{lal99} with the gas-grain chemical model of \cite{bl97}, we show that the
observed constant \NtwoHp\ abundance with extinction can be reproduced
provided that CO, which is a major destroyer of \NtwoHp , is 
depleting in the dense cores.  Since CO depletion is observed
this result is an important clue to the 
physical and chemical reason that allows \NtwoHp\ molecules to trace
the densest regions of molecular cores.   

(6) We also demonstrate that the observed \NtwoHp\ abundance threshold
($A_V^{th} \lesssim 4$ mag)
is partially due to the photodissociation of \Ntwo .  This highlights
an important application of this technique, which is the ability to 
test not only the chemical models of dense regions, and thereby
directly probe the gas-grain interaction, but also to provide a viable 
and completely new method for observationally testing models of
depth-dependent photodissociation rates.

\appendix
\section{Formation of  N$_2$H$^{+}$ Abundance}

The primary route to form N$_2$H$^{+}$ is through the following reaction:

\begin{equation}
\eqnum{A1}
\rm{H_3^+  +   N_2   \rightarrow   N_2H^+   +   H_2}.
\end{equation}

\noindent \Hthreep\ is the produced via cosmic ray ionization of 
H$_2$.  The main destruction pathways are 
dissociative electron recombination and a reaction with  CO (reactions
with O and C are of lesser import).
In equilibrium, the following expression can be derived for
the concentration of N$_2$H$^{+}$,

\begin{equation}
\eqnum{A2}
\rm{
n(N_2H^+) = \frac{n(H_3^+)n(N_2)k_{H_3^+,N_2}}{n(e^-)\alpha(N_2H^+) 
+ n(CO)k_{CO,N_2H^+}}
}.
\end{equation}

\noindent 
Where $\alpha(\rm{N_2H^+})$ is the dissociative recombination coefficient
and k$_{\rm{x,N_2H^+}}$ are the various reaction rates given
by \citet{mfw97}.  See also \citet{wzw92}. 
%This expression can be then solved for the abundance of N$_2$, which is
%
%\begin{equation}
%\eqnum{A3}
%\rm{
%n(N_2) = \frac{n(N_2H^+)}{n(H_3^+)k_{H_3^+,N_2}}
%\{n(e^-)\alpha(N_2H^+) + n(CO)k_{CO,N_2H^+}}.
%}
%\end{equation}

\newpage

\begin{figure}
\plotone{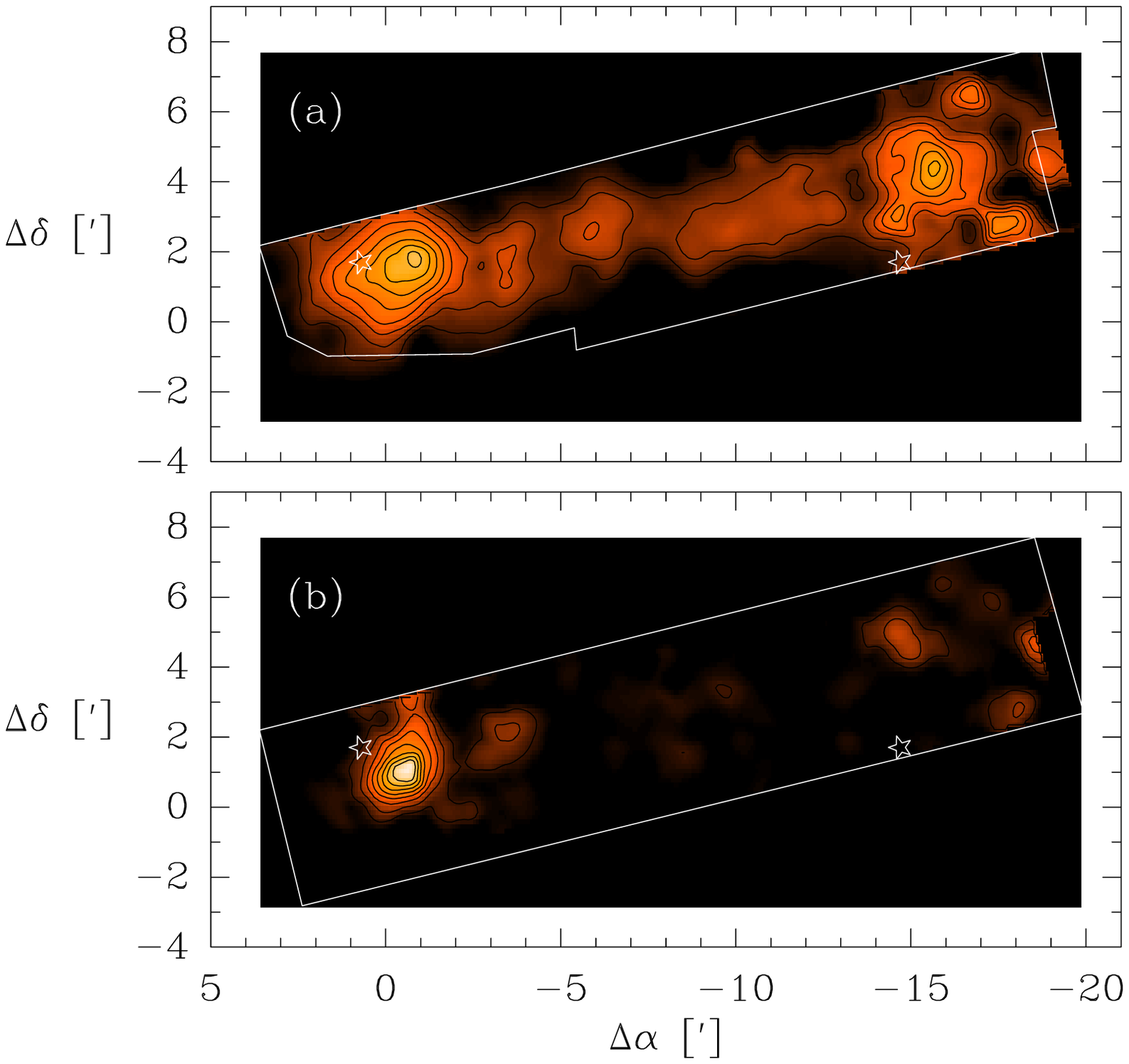}
\caption{
(a) Map of C$^{18}$O J = 1 $\rightarrow$ 0 integrated intensity in IC 5146.  Contour levels begin
with 0.3 K-km/s and are spaced by 0.2 K-km/s.  (b) Map of integrated 
N$_2$H$^+$ J = 2 $\rightarrow$ 1 emission.  Contour levels begin at 0.6 K-km/s and increase
in steps of 0.6 K-km/s.
The stars show the positions of IRAS point sources.
}
\label{fig_c18o_n2hp}
\end{figure}

\begin{figure}
\epsscale{0.8}
\plotone{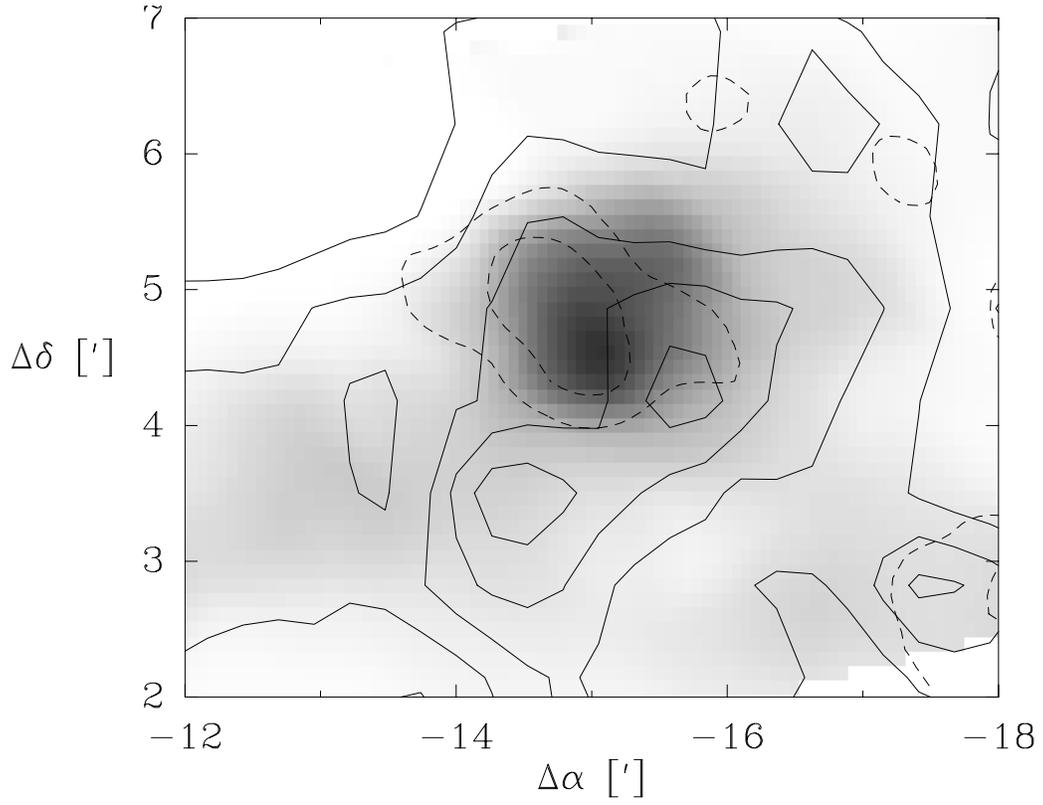}
\caption{
Comparison of C$^{18}$O J = 1 $\rightarrow$ 0  integrated emission distribution 
(solid contours), N$_2$H$^+$ J = 1 $\rightarrow$ 0 (dashed contours), with 
visual extinction (grey scale) towards the western core of the IC 5146 northern 
streamer.  Note how the N$_2$H$^+$ emission maximum corresponds with the peak
of visual extinction.  For C$^{18}$O the contour levels begin with  
and are spaced by 0.3 K-km/s, while for N$_2$H$^+$ contour levels begin with
and are spaced by 0.6 K-km/s. }
\label{fig_wc}
\end{figure}

\begin{figure}
\plotone{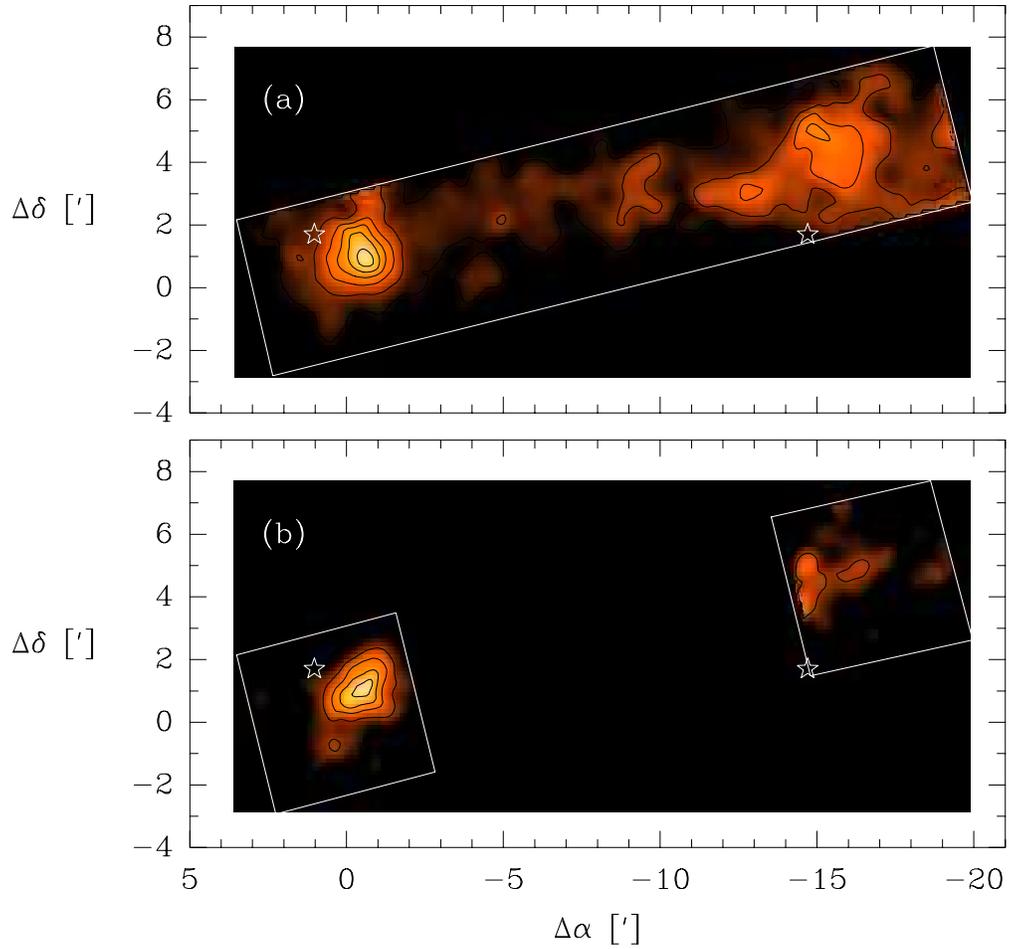}
\caption{
(a) Map of CS J = 1 $\rightarrow$ 0 integrated intensity in IC 5146.  Contour levels begin
with 0.5 K-km/s and are spaced by 0.5 K-km/s.  (b) Smaller maps of the integrated 
C$^{34}$S J = 2 $\rightarrow$ 1 emission.  Contour levels begin at 0.2 K-km/s and increase
in steps of 0.1 K-km/s.
The stars show the positions of IRAS point sources.
}
\label{fig_csdist}
\end{figure}

\begin{figure}
\plotone{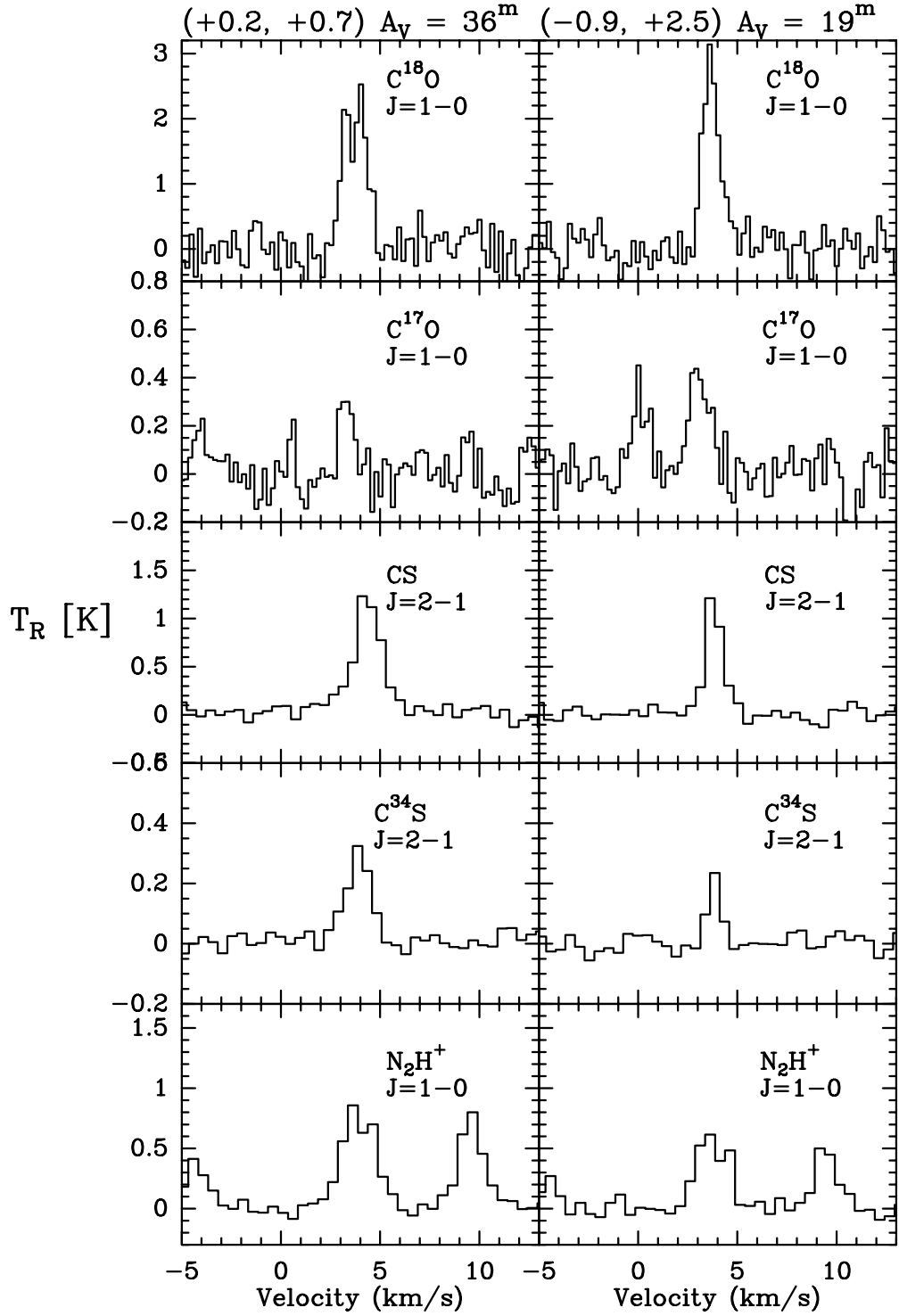}
\caption{  Plot showing selected spectra at two different positions in the Eastern core:  
($\Delta \alpha = +0\farcm 2$, $\Delta \delta = +0\farcm 7$) with \av\ = 36 mag and 
($\Delta \alpha = -0\farcm 9$, $\Delta \delta = +2\farcm 5$) with \av\ = 19 mag.  
}
\label{fig_spec}
\end{figure}

\begin{figure}
\epsscale{0.8}
\plotone{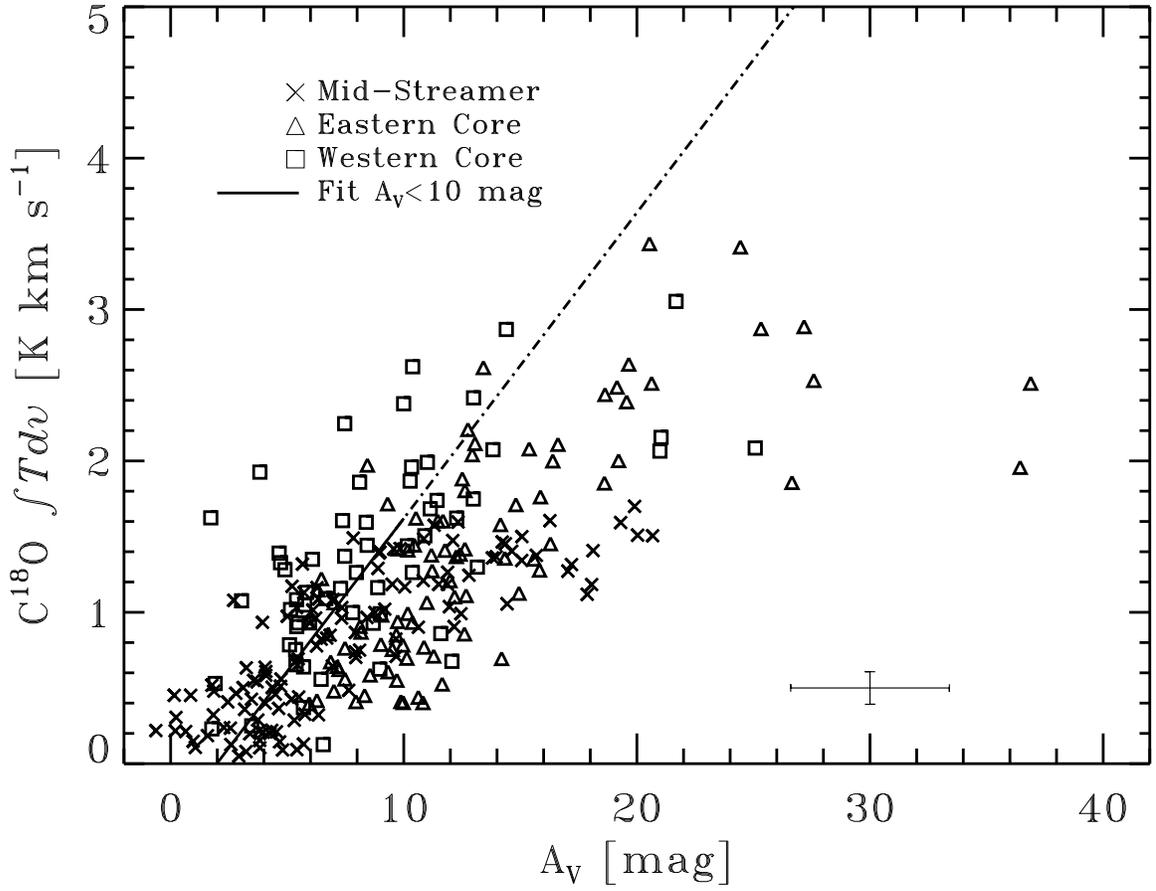}
\caption{
C$^{18}$O integrated intensity vs. visual extinction for the
entire Northern Streamer.  The best fit linear relationship for $A_V \leq
10$ mag is given as the solid line, and is extrapolated as the
dash-dot line  for $A_V > 10$ mag.  The median uncertainties in the
measurements are shown in the lower-right corner.  The eastern core,
western core, and mid-streamer are plotted as different symbols as
indicated by the legend.}
\label{fig_c18oint}
\end{figure}

\begin{figure}
\epsscale{0.8}
\plotone{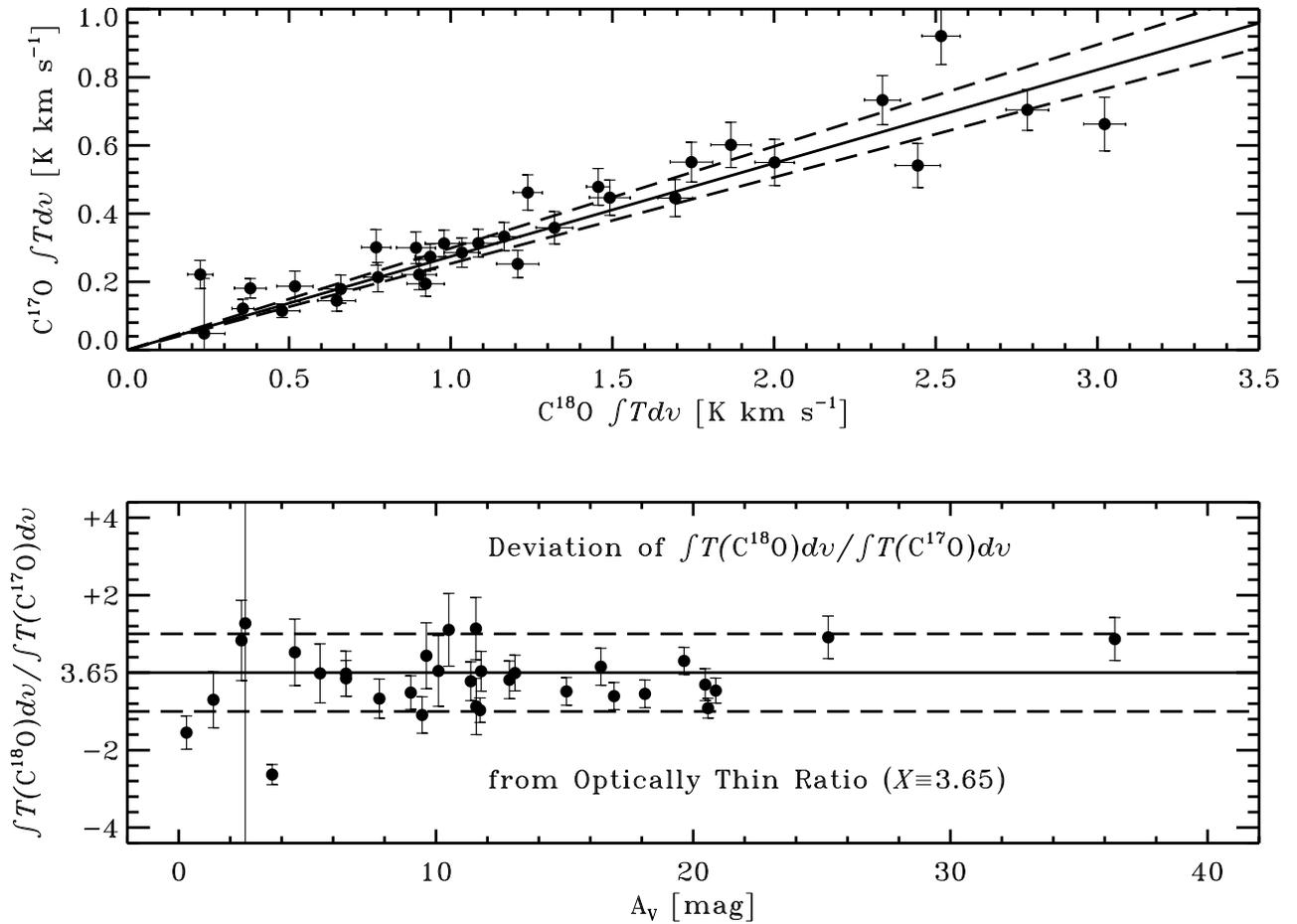}
\caption{
Comparison of the C$^{17}$O and C$^{18}$O integrated
intensities in the eastern core.  The solid line represents the expected
trend ($\pm 1\sigma$ -- dashed lines) for optically thin emission.  (b)
C$^{18}$O--C$^{17}$O ratio of integrated intensities vs. visual
extinction.  The solid line represents the expected ratio for optically
thin emission.  The dashed lines represent the median $1\sigma$
uncertainty of the measured ratios.}
\label{fig_c17o}
\end{figure}

\begin{figure}
\epsscale{0.8}
\plotone{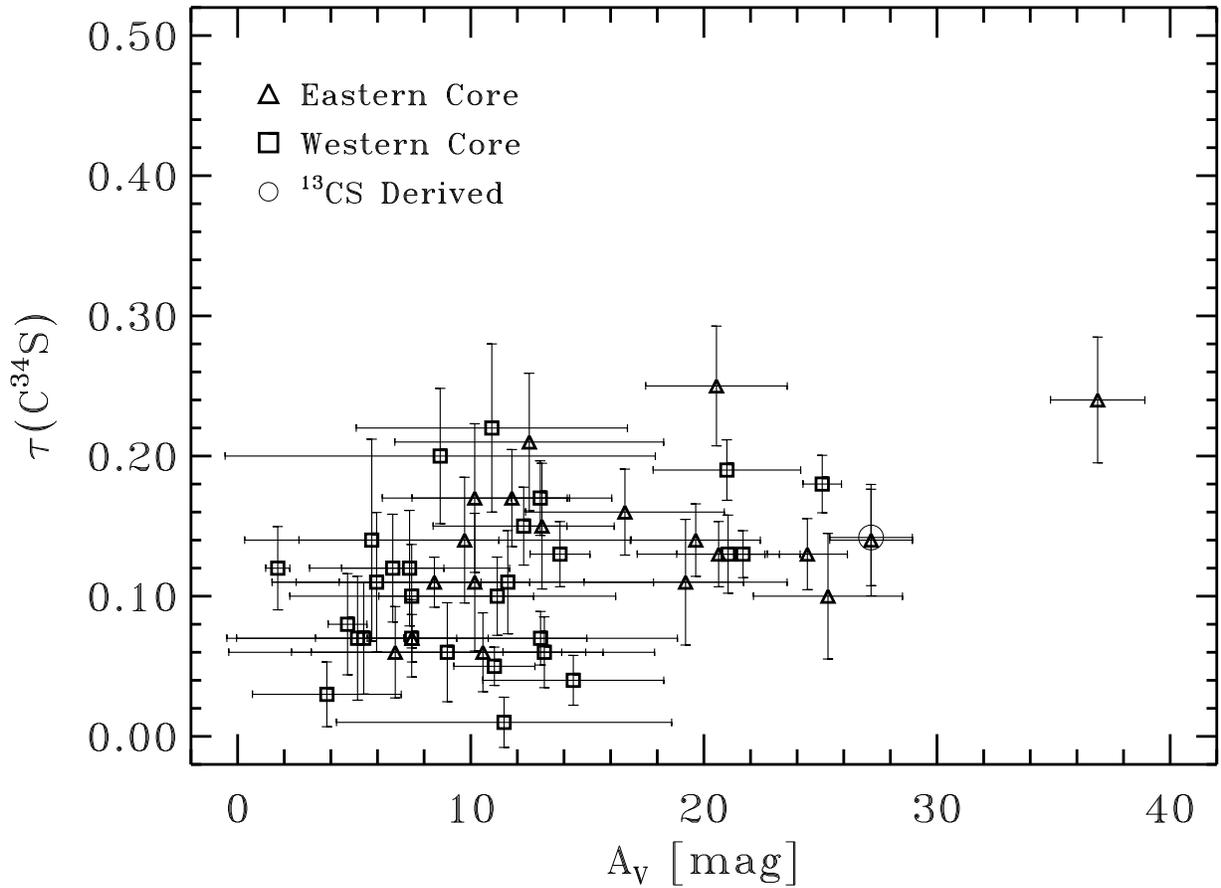}
\caption{
C$^{34}$S optical depth vs. visual extinction for the eastern
and western cores.  The eastern core and western core are
plotted as different symbols as indicated by the legend.  The {\em circle}
($\circ$) at $A_V \approx 27$ mag represents the optical depth as derived
from the $^{13}$CS data.  The C$^{34}$S optical depth measured for that
same line of sight overlaps the $^{13}$CS derived optical depth.}
\label{fig_cstau}
\end{figure}

\begin{figure}
\epsscale{0.8}
\plotone{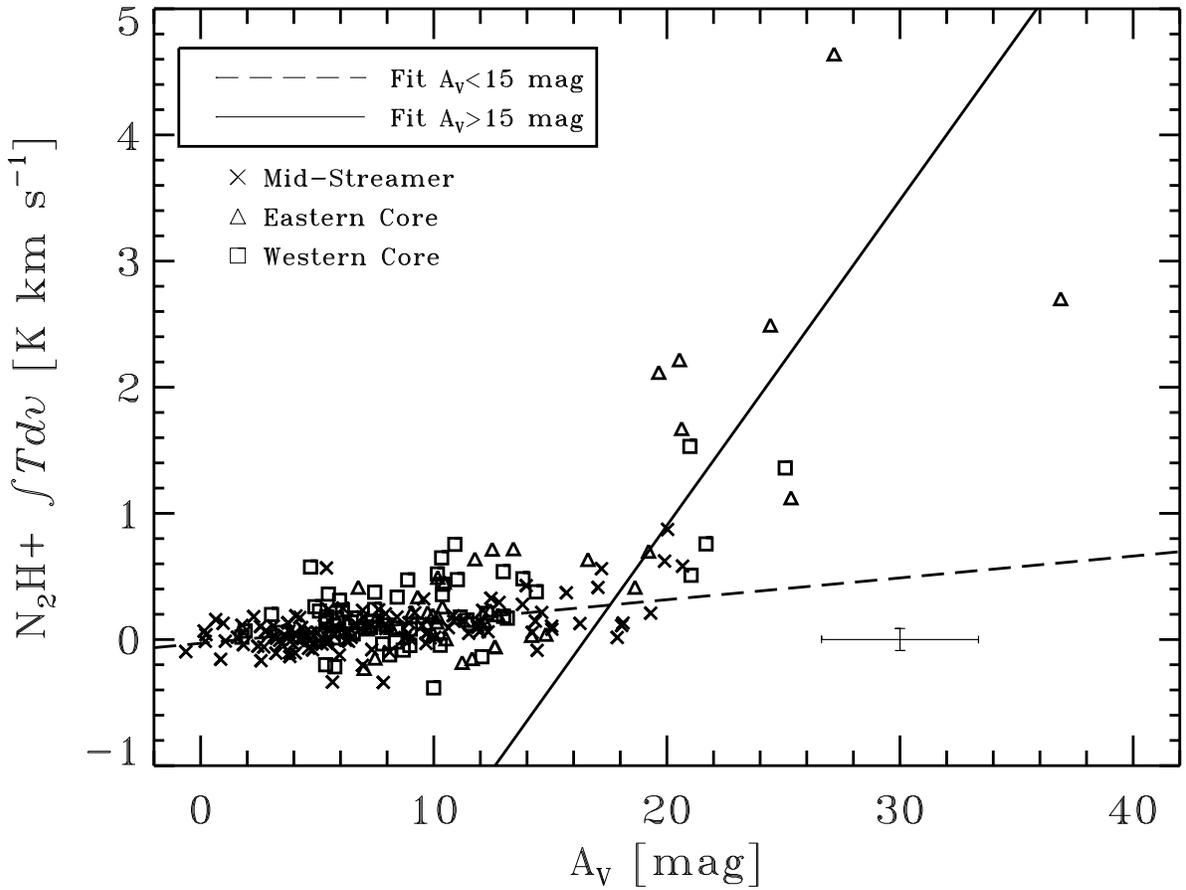}
\caption{
N$_2$H+ integrated intensity vs. visual extinction for the
entire Northern Streamer. The best fit linear relationship for $A_V < 15$
mag is plotted as the dashed line, and the best fit linear relationship
for $A_V > 15$ mag is plotted as the solid line.  The median uncertainties
in the measurements are shown in the lower-right corner.  The eastern
core, western core, and mid-streamer are plotted as different symbols as
indicated by the legend.}
\label{fig_n2hpint}
\end{figure}

\begin{figure}
%\epsscale{0.8}
\plotone{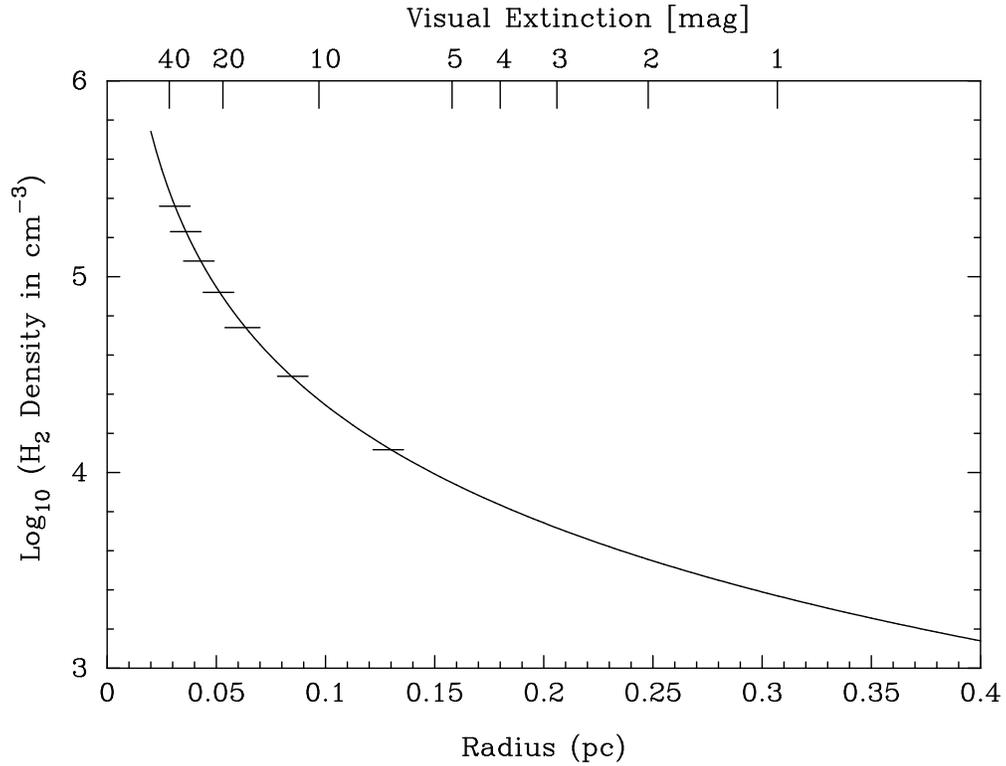}
\caption{
Volume density of molecular hydrogen as a function of visual extinction and radius.  This
profile is in good agreement with the profile of extinction with cloud radius determined in  
\citet{lal99}.  The horizontal hash marks found along the line representing the density
profile are the average densities used in the molecular abundance analysis for respective
bins of 5 mag starting with \av\ = 5 -- 10 mag.
}
\label{fig_denpro}
\end{figure}

\begin{figure}
%\epsscale{0.8}
\plotone{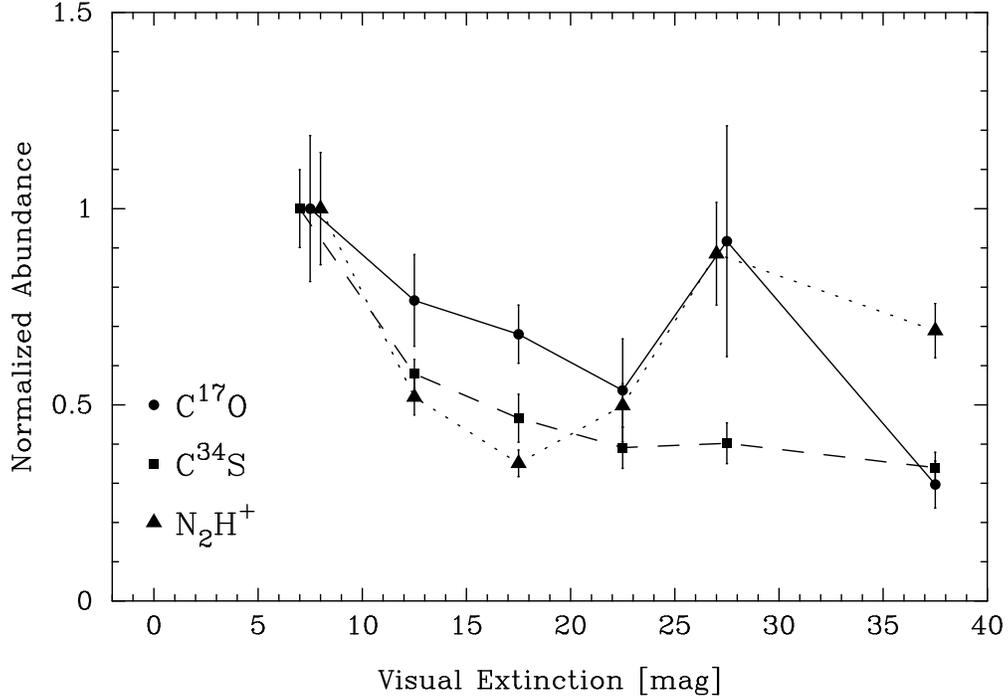}
\caption{
Normalized molecular abundances derived from combined molecular column densities and NIR
extinction measurements of C$^{17}$O (circles and solid line), \CtfS\ (squares and
dashed line), and
\NtwoHp\ (triangles and dotted line) relative to H$_2$ shown as a function of extinction in
the visual.  
The data are weighted averages
within bins of 5 mag, thus the errors are reflective of the total number of
data points included in the average and of the percentage error of the
individual data points.   The abundances are normalized using the data in the first bin
(A$_V$ = 5 -- 10 mag) in order to see the trend in relative abundances as a function
of extinction. 
}
\label{fig_abun}
\end{figure}

\begin{figure}
\epsscale{0.8}
\plotone{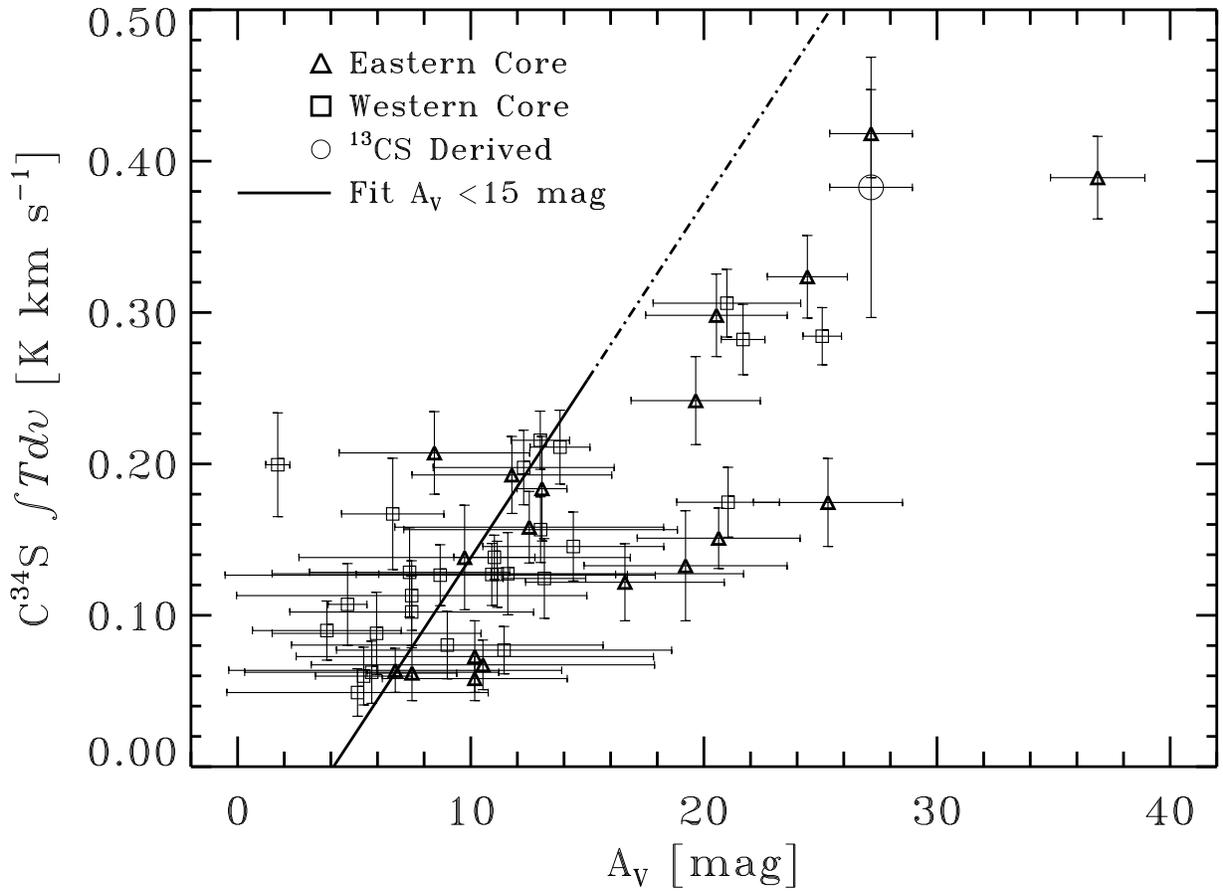}
\caption{
C$^{34}$S integrated intensity vs. visual extinction for the
eastern and western cores.  The best fit linear relationship for $A_V \leq
15$ mag is overplotted as a solid line, and is extrapolated as the
dash-dot line  for $A_V > 15$ mag.  The eastern core, western core, and
mid-streamer are plotted as different symbols as indicated by the legend. 
The {\em circle} ($\circ$) at $A_V \approx 27$ mag represents the
C$^{34}$S integrated intensity inferred from the measured $^{13}$CS
integrated intensity.  The C$^{34}$S integrated intensity measured for
that same line of sight overlaps the $^{13}$CS derived integrated
intensity.
}
\label{fig_c34sint}
\end{figure}

\begin{figure}
%\epsscale{0.8}
\plotone{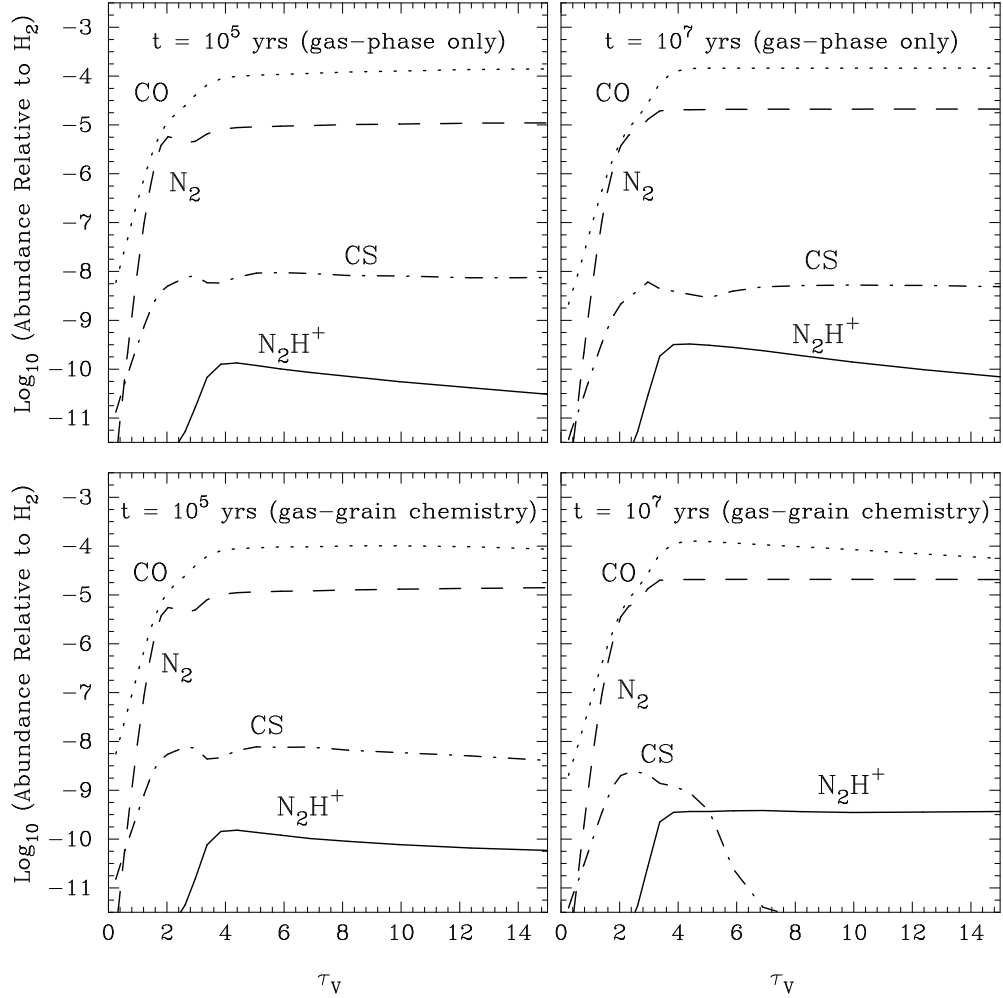}
\caption{
Profile of gas-phase abundances (abundances are relative to H$_2$) against
optical depth for surveyed species and \Ntwo .  The top panels are a
pure gas-phase chemical model, while the lower panels are a gas-grain chemical
model.  Similarly the left panels are for time (t) $= 10^5$ yr and 
the right panels t $= 10^7$ yr.  The physical conditions of the model
are given in \S 5.2. 
}
\label{fig_chemmod}
\end{figure}

%*********
%Table 3
%*********
\begin{deluxetable}{lccccc}
%\rotate
\tablewidth{0pt}
\tablecolumns{6}
\tablecaption{Summary of Molecular Line Observations}
\tablehead{

& \colhead{} & \colhead{Observation} &
\colhead{Mapping} & \colhead{Total $T_{int}$} & $T_{sys}$\\

\colhead{Map Region} & \colhead{Molecule} & \colhead{Date}
& \colhead{Mode} & \colhead{[sec]} & \colhead{[K]}

}
\startdata
\underline{Full Map}\tablenotemark{a} & C$^{18}$O & 1999 Mar 23 & PS\tablenotemark{b} & 450 & 300\\
& CS & 1999 Mar 23 & PS\tablenotemark{b} & 300 & 140\\
& N$_2$H+ & 1999 Mar 23, 25 & PS\tablenotemark{b} & 900\tablenotemark{c} & 170\\

\underline{East Core}\\
Full-Beam Footprint & C$^{34}$S & 1999 Mar 25 & PS\tablenotemark{b} & 1800 & 145\\
Single Array Pt.\tablenotemark{d} & $^{13}$CS & 1999 Mar 25 & PS\tablenotemark{b} & 2700 & 170\\
2 Single Array Pt.\tablenotemark{e}& C$^{18}$O & 2000 Jan 30 & FS & 600 & 180\\
2 Single Array Pt.\tablenotemark{e}& C$^{17}$O & 2000 Jan 30 & FS & 3600 & 300\\
Full-Beam Footprint\tablenotemark{f}& C$^{18}$O & 2000 Feb 01 & FS & 1200 & 180\\
\underline{West Core}\\
Full-Beam Footprint & C$^{34}$S & 2000 Jan 30 & FS & 2400 & 160\\

\enddata
\tablenotetext{a}{(0,0) position of map: $\alpha$ = 21:47:27; $\delta$=47:31:00 (J2000).}
\tablenotetext{b}{Reference position: $\alpha$ = 21:48:15; $\delta$=47:21:00 (J2000).}
\tablenotetext{c}{Central two footprints of N$_2$H+ map were observed for
a total integration time of 1500 sec.}
\tablenotetext{d}{Offset from map (0,0) = (0,0).}
\tablenotetext{e}{Offsets from map (0,0): Position 1 = (-2\farcm 69,1\farcm 4);
Position 2 = (-2\farcm 51, 2\farcm 15).}
\tablenotetext{f}{Offset from map (0,0) = (-0\farcm 369, -0\farcm 369).}
\end{deluxetable}
\clearpage
\begin{deluxetable}{rrrrrrr}
\tablenum{2}
\tablecolumns{7}
\tablewidth{5.0in}
\tablecaption{Average Molecular Abundances Relative to H$_2$\tablenotemark{a}}
\tablehead{
\colhead{A$_V$} &
\colhead{C$^{17}$O} &
\colhead{n$_{\rm C^{17}O}$} &
\colhead{C$^{34}$S} &
\colhead{n$_{\rm C^{34}S}$} &
\colhead{N$_2$H$^+$} &
\colhead{n$_{\rm N_2H^+}$}  \\
\colhead{(mag)} &
\colhead{($\times 10^{-7}$)} &
\colhead{} &
\colhead{($\times 10^{-11}$)} &
\colhead{} &
\colhead{($\times 10^{-11}$)} &
\colhead{}   
}
\startdata
 7.5 &   0.29$\pm$0.08  & 3 &    17.0$\pm$1.7  & 12 &    12.8$\pm$ 1.8  & 8\nl
12.5 &   0.27$\pm$0.04  & 5 &     9.8$\pm$0.6  & 17 &     6.7$\pm$0.6  & 18 \nl
17.5 &   0.26$\pm$0.03  & 4 &     7.9$\pm$1.0  & 3 &     4.5$\pm$0.4  & 9 \nl
22.5 &   0.20$\pm$0.05  & 3 &     6.6$\pm$0.9  & 6 &     6.4$\pm$0.7  & 8 \nl
27.5 &   0.34$\pm$0.11  & 1 &     6.8$\pm$0.9  & 3 &    11.3$\pm$1.7  & 3 \nl
37.5 &   0.09$\pm$0.02  & 1 &     5.8$\pm$0.7  & 1 &     8.8$\pm$0.9  & 1 \nl
\enddata
\tablenotetext{a}{n$_{\rm C^{18}O}$, n$_{\rm C^{34}S}$, n$_{\rm N_2H^+}$ are the number
of points included in binned average.}
\end{deluxetable}

\end{document}